\documentclass[trackchanges, twocolumn]{aastex701}

\newcommand{\oiii}{[O\,{\sc iii}]}

\newcommand{\nii}{[N\,{\sc ii}]}
\newcommand{\nev}{[Ne\,{\sc v}]}

\newcommand{\hei}{He\,{\sc i}}

\shorttitle{
Dust Around Little Red Dots
}
\shortauthors{Kiyota et al.}
\received{2026 August 11}
\revised{2026 October 8}
\accepted{2026 October 11}
\submitjournal{ApJ}

\begin{document}

\title{
ATLAS. III.\\
Dust Around Little Red Dots:\\
Hydrogen Line Ratios beyond Dust-free Non-Case B Models
}

\author[orcid=0009-0004-4332-9225]{Tomokazu Kiyota}
\affiliation{Astronomical Science Program, Graduate Institute for Advanced Studies, SOKENDAI, 2-21-1 Osawa, Mitaka, Tokyo 181-8588, Japan}
\affiliation{National Astronomical Observatory of Japan, 2-21-1 Osawa, Mitaka, Tokyo 181-8588, Japan}
\email[show]{tomokazu.kiyota@grad.nao.ac.jp} 
\correspondingauthor{Tomokazu Kiyota}

\author[orcid=0000-0002-1049-6658]{Masami Ouchi}
\affiliation{National Astronomical Observatory of Japan, 2-21-1 Osawa, Mitaka, Tokyo 181-8588, Japan}
\affiliation{Institute for Cosmic Ray Research, The University of Tokyo, 5-1-5 Kashiwanoha, Kashiwa, Chiba 277-8582, Japan}
\affiliation{Astronomical Science Program, Graduate Institute for Advanced Studies, SOKENDAI, 2-21-1 Osawa, Mitaka, Tokyo 181-8588, Japan}
\affiliation{Kavli Institute for the Physics and Mathematics of the Universe (WPI), University of Tokyo, Kashiwa, Chiba 277-8583, Japan}
\email{ouchims@icrr.u-tokyo.ac.jp}

\author[orcid=0009-0006-6763-4245]{Hiroto Yanagisawa}
\affiliation{Institute for Cosmic Ray Research, The University of Tokyo, 5-1-5 Kashiwanoha, Kashiwa, Chiba 277-8582, Japan}
\affiliation{Department of Physics, Graduate School of Science, The University of Tokyo, 7-3-1 Hongo, Bunkyo, Tokyo 113-0033, Japan}
\email[]{yana@icrr.u-tokyo.ac.jp}

\author[0000-0002-4225-4477]{Makoto Ando}
\affiliation{Institute for Cosmic Ray Research, The University of Tokyo, 5-1-5 Kashiwanoha, Kashiwa, Chiba 277-8582, Japan}
\email[]{mando@icrr.u-tokyo.ac.jp}

\author[0000-0002-6047-430X]{Yuichi Harikane}
\affiliation{Institute for Cosmic Ray Research, The University of Tokyo, 5-1-5 Kashiwanoha, Kashiwa, Chiba 277-8582, Japan}
\email[]{hari@icrr.u-tokyo.ac.jp}

\author[0009-0004-0381-7216]{Yuta Kageura}
\affiliation{Institute for Cosmic Ray Research, The University of Tokyo, 5-1-5 Kashiwanoha, Kashiwa, Chiba 277-8582, Japan}
\affiliation{Department of Physics, Graduate School of Science, The University of Tokyo, 7-3-1 Hongo, Bunkyo, Tokyo 113-0033, Japan}
\email[]{kageura@icrr.u-tokyo.ac.jp}

\author[0009-0000-1999-5472]{Minami Nakane}
\affiliation{Institute for Cosmic Ray Research, The University of Tokyo, 5-1-5 Kashiwanoha, Kashiwa, Chiba 277-8582, Japan}
\affiliation{Department of Physics, Graduate School of Science, The University of Tokyo, 7-3-1 Hongo, Bunkyo, Tokyo 113-0033, Japan}
\email[]{nakanem@icrr.u-tokyo.ac.jp} 

\author[0000-0002-0984-7713]{Yurina Nakazato}
\affiliation{Center for Computational Astrophysics, Flatiron Institute, 162 5th Avenue, New York, NY 10010}
\email[]{ynakazato@flatironinstitute.org}

\author[0000-0001-9011-7605]{Yoshiaki Ono}
\affiliation{Institute for Cosmic Ray Research, The University of Tokyo, 5-1-5 Kashiwanoha, Kashiwa, Chiba 277-8582, Japan}
\email[]{ono@icrr.u-tokyo.ac.jp}

\author[orcid=0009-0005-2897-002X]{Yui Takeda}
\affiliation{Astronomical Science Program, Graduate Institute for Advanced Studies, SOKENDAI, 2-21-1 Osawa, Mitaka, Tokyo 181-8588, Japan}
\affiliation{National Astronomical Observatory of Japan, 2-21-1 Osawa, Mitaka, Tokyo 181-8588, Japan}
\email[]{yui.takeda@grad.nao.ac.jp}

\begin{abstract}

We investigate broad hydrogen line ratios in little red dots (LRDs) using five high-redshift ($z>2$) sources from JWST/NIRSpec medium/high-resolution spectra in the DAWN JWST Archive and fifteen low-redshift sources ($z=0.1$--$0.9$) from the literature, all with broad H$\beta$ detected at $>5\sigma$. After carefully measuring the broad-line fluxes while accounting for absorption features and neighboring emission lines, we find that the broad H$\alpha$/H$\beta$ ratios are very high, ranging from 6 to 30, well above the Case B recombination value. 
Using plane-parallel \textsc{Cloudy} photoionization models with the total line emission from both sides of the slab,
we investigate the physical origin of the broad hydrogen line ratios beyond Case B, jointly modeling the Balmer and Paschen line ratios for the subset of one high-$z$ and two low-$z$ LRDs with detected broad Paschen lines. We find that one low-$z$ LRD is reproduced by a high gas number density ($\log (n_{\mathrm H}/{\rm cm^{-3}})\sim10$--11) on the broad-line H$\alpha$/H$\beta$--Pa$\beta$/Pa$\gamma$ plane, whereas the remaining two LRDs additionally require substantial dust extinction of $E(B-V)\gtrsim0.2$, even after accounting for non-Case B effects. Since the narrow H$\alpha$/H$\beta$ ratios do not indicate such large dust extinction, these results demonstrate that the obscuring dust is spatially associated with the broad-line region. Even without Paschen-line measurements, two and eight LRDs in the high- and low-redshift samples, respectively, exhibit H$\alpha$/H$\beta>13$, which cannot be reproduced by the non-Case B models, suggesting that dust obscuration might be common among LRDs, occurring in at least about half of the population. Such dust may represent a lower-column-density counterpart of the dusty torus in AGNs, reconciling the weak hot-dust emission.

\end{abstract}


\section{Introduction} \label{sec:intro} 

Little red dots (LRDs) are among the most striking discoveries enabled by the James Webb Space Telescope (JWST; \citealt{gardner23}). They are typically spatially unresolved in rest-frame optical JWST images and exhibit V-shaped spectral energy distributions (SEDs), with blue rest-frame ultraviolet (UV) continua, red rest-frame optical continua, and broad Balmer emission lines \citep[e.g.,][]{kocevski23, harikane23, matthee24, greene24, maiolino24, hviding25, jones25, labbe25, akins25, degraaff26-LRD, rusakov26}. 

Despite these broad lines, LRDs differ from classical active galactic nuclei (AGNs) in several respects. Many LRDs show weak X-ray emission \citep[e.g.,][]{ananna24, yue24, sacchi25, maiolino25}, a deficit of cold-dust emission relative to classical obscured AGNs or quasars (e.g., \citealt{casey25, setton25, leung25, ronayne26, xiao25}, see also \citealt{delvecchio25, brazzini26, barro26}), and weak variability \citep[e.g.,][]{kokubo25, zhang25, burke26}. These properties motivate models in which the physical conditions near the central engine differ from those in classical AGNs. 

One promising interpretation is that the accreting black hole is embedded in a dense gas \citep[e.g.,][]{naidu25, degraaff26-LRD, inayoshi25, inayoshi+maiolino25, matthee26, pacucci26, rusakov26, sneppen26, asada26, yanagisawa26, sun26-BHstar, sun26}. In this picture, the emergent spectrum is shaped by hydrogen radiative transfer in a dense, high-column-density gas, where line optical-depth effects become important. A key clue is that the broad Balmer decrement, especially broad H$\alpha$/H$\beta$, is often larger than that of the narrow component \citep[e.g.,][]{brooks25, nikopoulos25, chen26, geris26}. The narrow-line ratios are generally more consistent with Case B recombination and modest attenuation, whereas the enhanced broad-line decrement suggests that non-Case B processes in the nuclear broad-line gas may contribute. Dust attenuation may also contribute to the observed broad-line ratios \citep[e.g.,][]{brooks25, nikopoulos25, yan26, sneppen26, chang26}. 

Broad Balmer and Paschen lines provide valuable probes of this central engine. Dust attenuation, collisional excitation, and large line optical depths can modify the observed decrements relative to their Case B values. Balmer lines alone suffer from degeneracies between dust attenuation and non-Case-B physics \citep[e.g.,][]{kwan81, wills85}. Paschen lines are less sensitive to dust and therefore provide a crucial additional constraint. Joint analyses of broad Balmer and Paschen series can thus test whether the observed line ratios require dense, optically thick gas, dust attenuation, or both. 

Complementary evidence for dust is also provided by mid-infrared (MIR) emission. For example, stacked JWST Mid-Infrared Instrument (MIRI; \citealt{bouchet15, rieke15, wright23}) SEDs of high-redshift LRDs show an excess at $\lambda_{\rm rest}\sim3~\mu{\rm m}$, which is consistent with dust emission at $\sim1000$~K (\citealt{delvecchio25}, see also \citealt{brazzini26}). In addition, \citet{lin26} report the Wide-field Infrared Survey Explorer (WISE; \citealt{wright10}) detections of local LRDs at $z=0.1$--$0.2$ and infer warm/hot dust components from their SEDs. These MIR measurements are consistent with the presence of warm/hot dust, which might be located near the central engines. However, the MIR-emitting dust need not be identical to the material that attenuates the broad-line region along our line of sight. Constraining the dust location, geometry, and its connection to the line-emitting gas, therefore, requires analysis of hydrogen-line ratios. 

This paper is the third in the Archival and Theoretical study of LRDs with AGN comparison across Surveys (ATLAS) series. The ATLAS project investigates the LRD population using archival JWST spectroscopy, complemented by theoretical modeling and comparison samples of classical AGNs (\citealt{yanagisawa26, yanagisawa26_atlas2, kageura26}). 
In this paper, we jointly use broad Balmer and Paschen line ratios to assess whether dense, optically thick gas can reproduce the observed decrements without dust attenuation and, where it cannot, to constrain the additional attenuation required within the adopted model framework. We then discuss whether the inferred attenuation is consistent with available MIR constraints. Section~\ref{sec:data} describes the sample and the data. Section~\ref{sec:analysis} presents emission-line fitting and photoionization modeling. Section~\ref{sec:results} presents the measured line ratios, and Section~\ref{sec:discussion} discusses the implications for the central engines and dust geometry of LRDs. We summarize our conclusions in Section~\ref{sec:summary}. 
In this paper, we assume a flat $\Lambda$CDM cosmology with $H_0 = 67.7~\mathrm{km~s^{-1}~Mpc^{-1}}$, $\Omega_m = 0.31$, and $\Omega_\Lambda = 0.69$ \citep{planck20}. All magnitudes are in the AB system \citep{oke83}.

\section{Data} \label{sec:data}

\begin{figure}
    \includegraphics[width=1.0\linewidth]{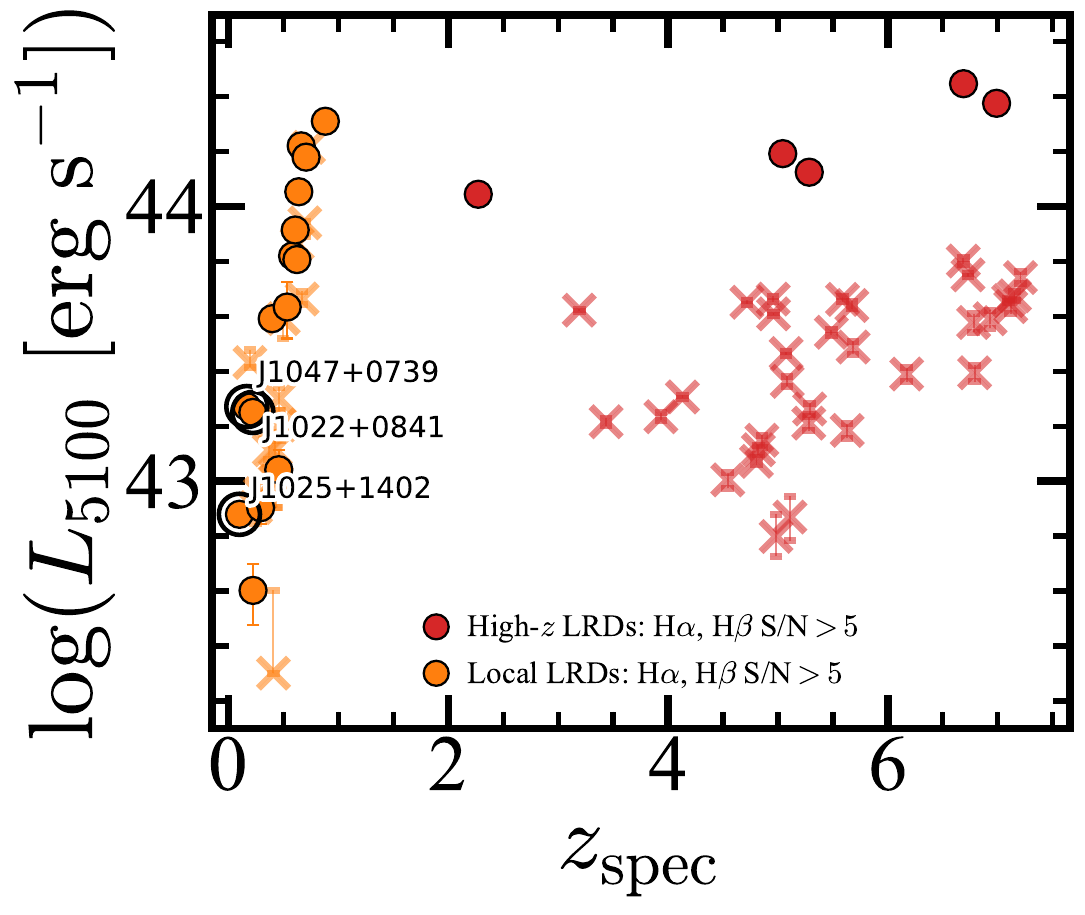}
    \caption{
    Relation between the rest-frame 5100\,\AA\ luminosity and the spectroscopic redshift for the primary sample and the auxiliary high-redshift lower-limit sources. The red and orange symbols show high- and low-redshift LRDs, respectively. The circles indicate sources with both H$\alpha$ and H$\beta$ detected at ${\rm S/N}\geq5$, while crosses indicate sources for which the H$\beta$ line has ${\rm S/N}<5$. The black circles and labels highlight the three local LRDs reported by \citet{lin26}. The $L_{5100}$ values are adopted from \citet{degraaff26-LRD}, \citet{lin26}, and \citet{lin26b}.
    }
\label{fig:L5100_z}
\end{figure}

We assemble a sample of 20 LRDs, comprising 15 low-redshift ($z=0.1$--$0.9$) and five high-redshift ($z=2.3$--$7.0$) sources. For the primary broad Balmer-decrement analysis, we require broad H$\beta$ to be detected at a signal-to-noise
ratio of ${\rm S/N}\geq5$. This criterion is applied to both the low- and high-redshift samples. The sample is selected from the literature and available spectroscopy and is not intended to be statistically complete. Its basic properties are summarized in Table~\ref{tab:sample} and Figure~\ref{fig:L5100_z}. 
The sample spans $\log(L_{\rm 5100}/{\rm erg~s^{-1}})=42.6$--$44.5$ as shown in Figure~\ref{fig:L5100_z}. 

We additionally identify high-redshift LRDs with suitable spectral coverage of both Balmer complexes but broad H$\beta$ detected at ${\rm S/N}<5$. These sources are excluded from the final sample and from the associated statistics because they provide only lower limits on broad H$\alpha$/H$\beta$. We show them separately in Figure~\ref{fig:L5100_z} and Figure~\ref{fig:broadHaHb_Haluminosity}. The emission-line fitting and lower-limit calculations are described in Section~\ref{subsec:flux-measurements}.

\begin{deluxetable*}{lcccccccccccc}
    \tablecaption{LRD sample in this study. \label{tab:sample}}
    \tabletypesize{\scriptsize}
    \tablewidth{0pt}
    \tablehead{
    \colhead{ID} & \colhead{R.A.} & \colhead{Decl.} & \colhead{Redshift} & \colhead{$L_{5100}$} & \multicolumn{5}{c}{Broad line} & \multicolumn{2}{c}{Narrow line} & \colhead{References} \\
    \cline{6-10} \cline{11-12}
    \colhead{} & \colhead{(deg)} & \colhead{(deg)} & \colhead{} & \colhead{} & \colhead{$\mathrm{H\alpha}$} & \colhead{$\mathrm{H\beta}$} & \colhead{$\mathrm{Pa\alpha}$} & \colhead{$\mathrm{Pa\beta}$} & \colhead{$\mathrm{Pa\gamma}$} & \colhead{$\mathrm{H\alpha}$} & \colhead{$\mathrm{H\beta}$} & \colhead{}
    }
    \colnumbers
    \startdata
    J102530.29+140207.3 & $156.37622$ & $14.03586$ & $0.101$ & $0.809_{-0.033}^{+0.033}$ & $87.6_{-2.56}^{+2.56}$ & $5.75_{-0.35}^{+0.35}$ & $22.7_{-0.26}^{+0.26}$ & $4.49_{-0.42}^{+0.42}$ & $3.03_{-0.13}^{+0.13}$ & $21.2_{-0.21}^{+0.21}$ & $6.20_{-0.05}^{+0.05}$ & L26a \\
    J104755.92+073951.2 & $161.98302$ & $7.66423$ & $0.168$ & $2.00_{-0.057}^{+0.057}$ & $293.6_{-6.19}^{+6.19}$ & $30.1_{-1.62}^{+1.62}$ & $88.4_{-0.59}^{+0.59}$ & $67.0_{-0.93}^{+0.93}$ & $30.6_{-0.43}^{+0.43}$ & $201.3_{-1.60}^{+1.60}$ & $62.1_{-0.52}^{+0.52}$ & L26a \\
    J102208.52+084156.1 & $155.53552$ & $8.69892$ & $0.223$ & $1.90_{-0.033}^{+0.033}$ & $554.4_{-13.7}^{+13.7}$ & $16.5_{-0.51}^{+0.51}$ & $125.0_{-3.05}^{+3.05}$ & $79.3_{-2.84}^{+2.84}$ & $21.0_{-1.16}^{+1.16}$ & $30.3_{-7.68}^{+7.68}$ & $15.4_{-0.21}^{+0.21}$ & L26a \\
    J132137.00-021417.04 & $200.40417$ & $-2.23807$ & $0.224$ & $0.4_{-0.1}^{+0.1}$ & $41.0_{-0.60}^{+0.60}$ & $2.60_{-0.40}^{+0.40}$ & \nodata & \nodata & \nodata & $13.0_{-0.30}^{+0.30}$ & $5.10_{-0.20}^{+0.20}$ & L26b \\
    J134317.81+393418.07 & $205.82421$ & $39.57169$ & $0.293$ & $0.8_{-0.1}^{+0.1}$ & $66.4_{-1.30}^{+1.50}$ & $5.50_{-0.80}^{+0.80}$ & \nodata & \nodata & \nodata & $54.8_{-0.70}^{+0.70}$ & $17.3_{-0.40}^{+0.40}$ & L26b \\
    J082921.37+131237.44 & $127.33904$ & $13.21040$ & $0.399$ & $3.9_{-0.2}^{+0.2}$ & $443.2_{-34.1}^{+62.6}$ & $20.4_{-1.90}^{+2.30}$ & \nodata & \nodata & \nodata & $180.7_{-2.00}^{+2.30}$ & $57.0_{-0.90}^{+1.10}$ & L26b \\
    J024337.99+034915.97 & $40.90829$ & $3.82110$ & $0.458$ & $1.1_{-0.2}^{+0.2}$ & $175.2_{-3.70}^{+3.70}$ & $11.6_{-1.80}^{+1.80}$ & \nodata & \nodata & \nodata & $53.1_{-1.50}^{+1.40}$ & $17.9_{-0.60}^{+0.60}$ & L26b \\
    J164102.65+070806.47 & $250.26104$ & $7.13513$ & $0.535$ & $4.3_{-1}^{+1}$ & $635.4_{-27.7}^{+32.4}$ & $30.1_{-3.10}^{+3.10}$ & \nodata & \nodata & \nodata & $57.3_{-12.2}^{+15.9}$ & $13.0_{-0.80}^{+0.80}$ & L26b \\
    J212725.88-044808.92 & $321.85783$ & $-4.80248$ & $0.584$ & $6.6_{-0.5}^{+0.6}$ & $485.5_{-36.0}^{+48.9}$ & $76.6_{-11.5}^{+13.6}$ & \nodata & \nodata & \nodata & $42.7_{-3.90}^{+4.80}$ & $13.2_{-1.10}^{+1.00}$ & L26b \\
    J104242.43+372147.63 & $160.67679$ & $37.36323$ & $0.608$ & $8.2_{-0.6}^{+0.6}$ & $1435_{-37.2}^{+34.8}$ & $105.1_{-8.00}^{+8.50}$ & \nodata & \nodata & \nodata & $69.0_{-20.8}^{+31.2}$ & $14.7_{-1.10}^{+1.10}$ & L26b \\
    J142337.59+520216.05 & $215.90662$ & $52.03779$ & $0.624$ & $6.4_{-0.3}^{+0.2}$ & $795.0_{-28.3}^{+21.4}$ & $72.5_{-10.0}^{+11.2}$ & \nodata & \nodata & \nodata & $45.7_{-2.40}^{+1.70}$ & $15.0_{-0.70}^{+0.60}$ & L26b \\
    J165450.36+033741.74 & $253.70983$ & $3.62826$ & $0.641$ & $11.3_{-0.5}^{+0.6}$ & $1731_{-68.4}^{+72.1}$ & $140.5_{-16.0}^{+23.3}$ & \nodata & \nodata & \nodata & $71.0_{-14.4}^{+14.1}$ & $19.2_{-1.30}^{+1.20}$ & L26b \\
    J094411.31-024908.65 & $146.04712$ & $-2.81907$ & $0.662$ & $16.6_{-0.5}^{+0.6}$ & $3262_{-122.9}^{+90.2}$ & $297.5_{-8.40}^{+10.1}$ & \nodata & \nodata & \nodata & $87.1_{-14.2}^{+12.4}$ & $18.8_{-1.70}^{+1.50}$ & L26b \\
    J164637.91+142648.62 & $251.65796$ & $14.44684$ & $0.707$ & $15.1_{-0.7}^{+0.4}$ & $2794_{-183.4}^{+117.3}$ & $259.3_{-17.9}^{+41.6}$ & \nodata & \nodata & \nodata & $182.5_{-11.0}^{+11.2}$ & $56.8_{-3.90}^{+2.20}$ & L26b \\
    J102553.75+502843.24 & $156.47396$ & $50.47868$ & $0.882$ & $20.4_{-1.5}^{+1.5}$ & $5850_{-75.9}^{+77.1}$ & $441.7_{-50.8}^{+87.2}$ & \nodata & \nodata & \nodata & $170.7_{-27.2}^{+24.9}$ & $32.8_{-2.60}^{+2.90}$ & L26b \\
    JADES-GN-28074 & $189.06459$ & $62.27382$ & $2.260$ & $11.6_{-0.0254}^{+0.0213}$ & $2782_{-27.2}^{+27.8}$ & $200.7_{-5.31}^{+6.29}$ & \nodata & $473.1_{-13.1}^{+10.8}$ & $208.3_{-8.9}^{+11.4}$ & $426.3_{-16.9}^{+13.2}$ & $93.1_{-3.67}^{+3.39}$ & This work, J24, dG25 \\
    JADES-GN-68797 & $189.22914$ & $62.14619$ & $5.039$ & $16.2_{-0.098}^{+0.120}$ & $6285_{-219.3}^{+205.6}$ & $403.6_{-43.9}^{+43.2}$ & \nodata & \nodata & \nodata & $533.6_{-144.9}^{+200.3}$ & $105.3_{-22.7}^{+19.8}$ & This work, dG25 \\
    RUBIES-EGS-42046 & $214.79537$ & $52.78885$ & $5.276$ & $13.9_{-0.120}^{+0.119}$ & $5552_{-263.1}^{+369.9}$ & $551.2_{-58.6}^{+54.8}$ & \nodata & \nodata & \nodata & \nodata & \nodata & This work, dG25 \\
    RUBIES-EGS-49140 & $214.89225$ & $52.87741$ & $6.685$ & $29.1_{-0.217}^{+0.217}$ & $6983_{-162.2}^{+137.6}$ & $825.6_{-35.3}^{+36.3}$ & \nodata & \nodata & \nodata & \nodata & \nodata & This work, dG25 \\
    RUBIES-EGS-55604 & $214.98303$ & $52.95600$ & $6.983$ & $24.7_{-0.209}^{+0.235}$ & $8370_{-210.6}^{+229.8}$ & $658.3_{-32.2}^{+30.1}$ & \nodata & \nodata & \nodata & $82.2_{-55.3}^{+65.2}$ & $24.0_{-12.3}^{+12.3}$ & This work, dG25 \\
    RUBIES-UDS-40579 & $34.24420$ & $-5.24587$ & $3.104$ & $14.3_{-0.0614}^{+0.0653}$ & $3021_{-178.4}^{+186.9}$ & \nodata & \nodata & $815.1_{-29.1}^{+30.2}$ & $556.4_{-97.6}^{+112.8}$ & $225.0_{-130.1}^{+142.4}$ & \nodata & This work, dG25, W25 \\
    \enddata
    \tablecomments{
    Primary LRD sample and a supplementary Paschen-line source. 
    For the primary LRD sample, sources with both broad H$\alpha$ and broad H$\beta$ detected with ${\rm S/N>5}$ are shown. $L_{5100} \equiv \lambda L_\lambda(5100\text{\AA})$ is in units of $10^{43}\,\mathrm{erg\,s^{-1}}$. The line luminosities are in units of $10^{40}\,\mathrm{erg\,s^{-1}}$. The line luminosities of the low-redshift LRDs are taken from \citet{lin26} and \citet{lin26b}. For the high-redshift LRDs whose narrow H$\alpha$ and H$\beta$ are not constrained in the fitting, we show no measurements. 
    As a reference, RUBIES-UDS-40579 is included separately as a supplementary source with multiple Paschen lines but without H$\beta$ coverage. 
    All luminosities and their uncertainties, including values taken from the literature, are rescaled to the cosmology adopted in this work. 
    L26a: \citet{lin26}; L26b: \citet{lin26b}; J24: \citet{juodvbalis24}; W25: \citet{wang25}; dG25: \citet{degraaff25}.
    }
\end{deluxetable*}

%%% fig: L5100 vs. z %%%
%%%%%%%%%%%%%%%%%%%%%%%%%

\subsection{Low-redshift LRDs} \label{subsec:local-sample}

The low-redshift sample consists of 15 sources: three LRDs reported by \citet{lin26} and 12 LRDs from \citet{lin26b}. \citet{lin26} report three local LRDs: J102530.29+140207.3, J104755.92+073951.2, and J102208.52+084156.1. Hereafter, we refer to these objects as J1025+1402, J1047+0739, and J1022+0841, respectively. \citet{lin26} identified them as local analogs of high-redshift LRDs based on their V-shaped spectral energy distributions (SEDs) and broad Balmer emission lines from the Sloan Digital Sky Survey (SDSS; \citealt{york00}) data. 
The local LRD identifications are based on the combination of a V-shaped UV--optical continuum, compact rest-frame optical morphology, and broad Balmer emission lines. For the three SDSS sources, \citet{lin26} searched for GALEX+SDSS SEDs resembling redshifted high-redshift LRD templates, applied a compactness criterion, and assessed their continua and emission-line properties.
J1025+1402 was also discussed by \citet{ji26}. We adopt the broad Balmer and Paschen line fluxes reported by \citet{lin26}. 

These three local LRDs also have mid-infrared photometry from the WISE W1, W2, W3, and W4 bands. The WISE fluxes are summarized in Appendix~\ref{app:wise-photometry}.
\citet{lin26} modeled this photometry using fixed-temperature dust templates \citep{lyu21} at approximately 90, 300, and 1000~K, together with an additional $\sim2000$~K component where required. We use these warm/hot-dust constraints in Section~\ref{sec:discussion} when discussing the dust properties of LRDs. 

In addition, \citet{lin26b} report a larger sample of local LRDs at $z=0.2$--$0.9$ (27 sources) from the Dark Energy Spectroscopic Instrument \citep[DESI;][]{desi24, desi26} data release 1. The DESI search of \citet{lin26b} selected sources with red optical and blue UV continua, PSF-like morphologies, and distinct broad Balmer components. It also applied additional criteria involving a declining near-infrared continuum, weak \nii\ emission, and the absence of \nev\ emission. These selections target the characteristic observational properties of high-redshift LRDs.
From these, we use the sources with both broad H$\alpha$ and H$\beta$ emission lines detected with ${\rm S/N}\geq5$, and the resulting number of sources is 12 from \citet{lin26b}. We utilize the broad Balmer line fluxes of these objects reported in \citet{lin26b}.

\subsection{High-redshift LRDs} \label{subsec:highz-sample} 

We construct the high-redshift sample from the LRD compilation of \citet{degraaff26-LRD}, which is based on version 4.4 of the DAWN JWST Archive (DJA; \citealt{heintz24, degraaff25_dja, valentino25, brammer_2025_15472354})\footnote{\url{https://dawn-cph.github.io/dja/}}. All spectra in the DJA are reduced with \texttt{msaexp} \citep{brammer23}. In brief, \citet{degraaff26-LRD} have selected sources with V-shaped SEDs in JWST/Near Infrared Spectrograph (NIRSpec; \citealt{jakobsen22}) PRISM spectra and compact morphology in JWST/Near Infrared Camera (NIRCam; \citealt{rieke23}) F444W imaging. 

From this compilation, we retain only sources with available JWST/NIRSpec medium ($R\sim1000$) or high ($R\sim2700$) resolution grating spectra. For measurements of the H$\alpha$/H$\beta$ ratio, we require both Balmer lines to fall within the wavelength coverage of the grating spectra. 
For each source, we use the catalog redshift to compute the expected observed-frame line centers, and require the spectra to cover a velocity interval of $\pm1000~{\rm km~s^{-1}}$ around both H$\alpha$ and H$\beta$ emission lines. This selection yields 37 sources at $z=2$--$7$.

For the Paschen line analysis, we additionally identify two objects with multiple Paschen lines in the grating spectra: Rosetta Stone at $z=2.26$ (JADES-GN-28074; \citealt{juodvbalis24, brazzini25}) and RUBIES-BLAGN-1 (catalog ID RUBIES-UDS-40579) at $z=3.10$ \citep{wang25}. 
The Rosetta Stone is already included in the H$\alpha$/H$\beta$-selected sample, while RUBIES-BLAGN-1 is only used specifically for the Paschen-line analysis. 
For the 37 selected sources, we downloaded the publicly available JWST/NIRSpec grating spectra from DJA version 4.4 and used these spectra for the emission-line measurements described in Section~\ref{subsec:flux-measurements}. The primary high-redshift sample therefore comprises five sources with both broad H$\alpha$ and broad H$\beta$ detected at ${\rm S/N}\geq5$. RUBIES-UDS-40579 is included only as a supplementary Paschen-line source and is not counted among the 20 primary-sample LRDs (Table~\ref{tab:sample}).

\section{Analysis} \label{sec:analysis} 

\subsection{Emission-line Fitting and Flux Measurements} \label{subsec:flux-measurements}

%%% fig: Fitting example %%%%%%
\begin{figure*}
    \includegraphics[width=1.0\linewidth]{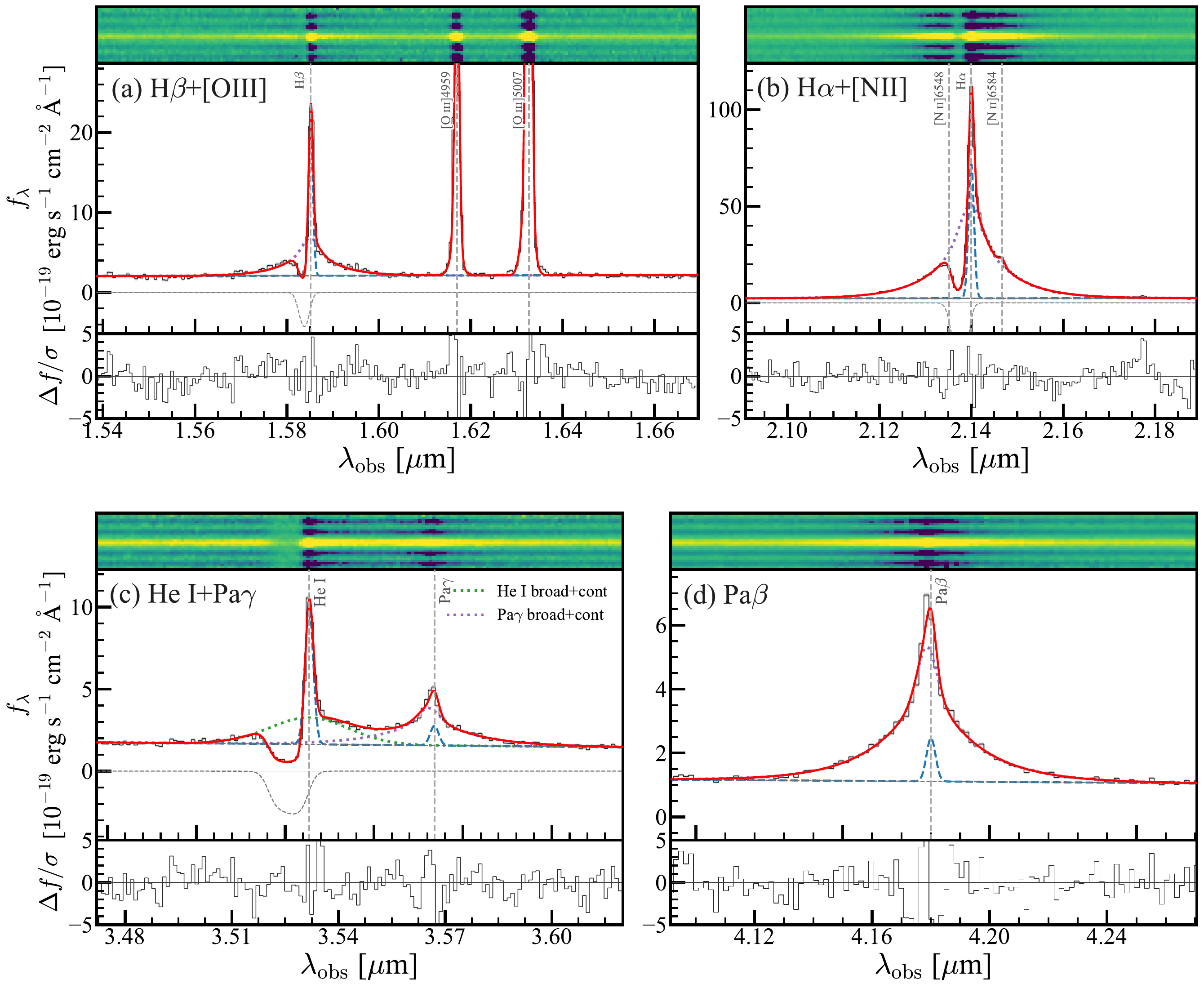}
    \caption{ 
    Example of the spectral fitting for JADES-GN-28074
    (Rosetta Stone).
    The JWST/NIRSpec spectra cover
    a) H$\beta$+\oiii$\lambda\lambda4959,5007$ (upper left),
    b) H$\alpha$+\nii$\lambda\lambda6548,6584$ (upper right),
    c) \hei~$\lambda10830$+Pa$\gamma$ (lower left), and
    d) Pa$\beta$ (lower right).
    Within each panel, the upper, middle, and lower sub-panels show the two-dimensional spectrum, the one-dimensional spectrum with the fitted model, and the residuals $(f_{\mathrm{obs}}-f_{\mathrm{model}})/\sigma$, respectively.
    The black histogram and gray-shaded region show the observed spectrum and its $1\sigma$ uncertainties, respectively. 
    The red solid curve shows the adopted spectral model evaluated at the posterior-median parameters. 
    Blue dashed curves show the narrow permitted-line emission, purple dotted curves show the broad Balmer and Paschen components, and the green dotted curve in
    panel~(c) shows the intrinsic broad \hei\ component.
    These emission components are displayed with the fitted
    continuum added. 
    Gray dashed curves show the continuum and, where applicable, the absorption.
    Vertical gray dashed lines mark the wavelengths of the labeled transitions at the fitted narrow-line redshift.
    Wavelengths are shown in the observed frame.
    }
\label{fig:fitting}
\end{figure*}
%%%%%%%%%%%%%%%%%%%%%%%%%%%%%%

We measure the Balmer and Paschen emission-line fluxes in the high-redshift sample. For each spectral region, we first determine the maximum-likelihood solution by minimizing $\chi^2$ with the Levenberg--Marquardt algorithm implemented in \texttt{lmfit} (version 1.3.4; \citealt{newville25}). We then use the best-fitting parameters to initialize Markov chain Monte Carlo (MCMC) sampling with \texttt{emcee} \citep{mackey13}. We adopt a Gaussian likelihood based on the flux-density uncertainties. The H$\beta$+\oiii\ complex, the H$\alpha$+\nii\ complex, and each Paschen-line region are fitted separately. All rest-frame wavelengths used in the emission-line fitting are in vacuum (e.g., 5008.240\,\AA\ for \oiii$\lambda5007$), although we follow the long-standing nomenclature based on their air wavelengths. 

For the H$\beta$+\oiii$\lambda\lambda4959, 5007$ complex, our baseline model contains narrow Gaussian components for H$\beta$ and \oiii$\lambda\lambda4959,5007$, a broad H$\beta$ component modeled with an exponential profile, and a linear continuum. 
The narrow H$\beta$ and \oiii\ components are constrained to have the same velocity widths and redshifts, while their fluxes are allowed to vary independently. 
The \oiii\ doublet flux ratio is fixed to 
$f_{\rm [O\,III]\lambda5007}/f_{\rm [O\,III]\lambda4959}=2.98$ \citep{storey00}. 

We also consider an additional Gaussian outflow component for \oiii\ emission lines. We define the difference in the Bayesian Information Criterion (BIC; \citealt{schwarz78}) as $\Delta{\rm BIC}={\rm BIC}_{\rm no\,outflow}-{\rm BIC}_{\rm outflow}$, where ${\rm BIC}_{\rm no\,outflow}$ (${\rm BIC}_{\rm outflow}$) is the BIC without (with) the outflow component, which shares redshift, velocity width,
and their flux ratio to be fixed to the same theoretical value. We include the outflow component when $\Delta\mathrm{BIC}>15$. In these cases, we also include an H$\beta$ outflow component whose velocity center and width are tied to those of the \oiii\ outflow component, while its flux is allowed to vary freely. 

We fit the H$\alpha$+\nii\ complex with narrow Gaussian components for H$\alpha$ and \nii$\lambda\lambda6548,6584$, a broad exponential H$\alpha$ component \citep[e.g.,][]{scholtz26}, and a linear continuum. 
The narrow components share a redshift and velocity width. 
The \nii\ doublet flux ratio is fixed to $f_{\rm [N\,II]\lambda6584}/f_{\rm [N\,II]\lambda6548}=3.05$ \citep{storey00}. 
The H$\alpha$ outflow component is also included using the same procedure as for the H$\beta$+\oiii, and its kinematics are tied to those measured from \oiii\ emission lines. 

We also perform fits to the H$\alpha$+\nii\ complex, including an absorption component described by a partial-covering absorption model,
\begin{equation}
    T(\lambda) = 1 - C_f + C_f e^{-\tau(\lambda)},
\end{equation}
where $C_f$ is the covering factor and $\tau(\lambda)$ is the optical-depth profile \citep[e.g.,][]{scholtz26}. 
This absorption component attenuates the broad component and continuum. 
We model the optical-depth profile as a Gaussian in velocity space,
\begin{equation}
    \tau(\lambda) = \tau_0 
    \exp\left[
    -\frac{1}{2}
    \left(
    \frac{v(\lambda)-\Delta v}{\sigma_v}
    \right)^2
    \right],
\end{equation}
where $\tau_0$ is the central optical depth, $\Delta v$ is the absorber velocity shift from the systemic redshift, and $\sigma_v$ is its velocity dispersion. 
The absorption component is included when the absorption model is favored over the no-absorption model by $\Delta{\rm BIC}>15$. 
The same procedures are independently applied to the other Balmer (H$\beta$) and Paschen series to measure absorption features. 

We model each covered Paschen region with a narrow Gaussian component, a broad exponential component, and a local linear continuum. An absorption component is added when favored by the same $\Delta{\rm BIC}>15$ criterion. 
Because broad \hei$\lambda10830$ overlaps Pa$\gamma$, the two lines are fitted simultaneously with separate broad exponential components.

The full model is convolved with the wavelength-dependent line-spread function (LSF). 
We adopt an effective resolving power of $R_{\rm eff}(\lambda)=1.8 \times R_{\rm doc}(\lambda)$, where $R_{\rm doc}(\lambda)$ is the wavelength-dependent resolving power from the JWST/NIRSpec documentation\footnote{\url{https://jwst-docs.stsci.edu/jwst-near-infrared-spectrograph/nirspec-instrumentation/nirspec-dispersers-and-filters}}, to account for the compact morphology of LRDs \citep{degraaff24}. 
This corresponds to an instrumental full width at half maximum (FWHM) of 
${\rm FWHM}_{\rm inst}(\lambda)=\lambda/R_{\rm eff}(\lambda)$. 
The LSF is assumed to be Gaussian in the fitting. 
Figure~\ref{fig:fitting} shows examples of the fitting results for the H$\beta$+\oiii, H$\alpha$+\nii, \hei$\lambda10830$+Pa$\gamma$ complex, and Pa$\beta$ of JADES-GN-28074 (Rosetta Stone). 

We report the median posterior line fluxes, with uncertainties given by the 16th and 84th percentiles of the posterior distributions. 
For objects in which the broad H$\beta$ component was not detected at ${\rm S/N}\geq5$, we refitted the H$\beta$ using the broad H$\alpha$ profile as a template. 
Specifically, exponential width and redshift are fixed at those of H$\alpha$, while the H$\beta$ normalization was allowed to vary. 
The uncertainty in the H$\alpha$ broad-line profile is propagated in the H$\beta$ MCMC approach. 
We define the 3$\sigma$ upper limit on the broad H$\beta$ flux as the 99.865th percentile of the posterior flux distribution. 
We calculate line-ratio posteriors from the relevant line-flux posteriors.

\subsection{Cloudy Modeling} \label{subsec:cloudy}

We model the hydrogen-line ratios with photoionization calculations using \textsc{Cloudy} version c23.01 \citep{ferland98, Gunasekera23}. 
The purpose of this modeling is to identify the range of physical conditions capable of reproducing the observed Balmer and Paschen line ratios. 
The models are not intended to determine a unique gas density, column density, or ionization parameter for each source. Instead, they provide a feasibility grid to assess whether the observed broad-line ratios can be reproduced by high-density and optically thick gas without dust attenuation.

We use constant-density, plane-parallel slab models over the $\Phi(\mathrm{H})$--$n_{\rm H}$ plane, where $\Phi(\mathrm{H})$ is the incident ionizing photon flux and $n_{\rm H}$ is the hydrogen number density (e.g., \citealt{korista04, Schnorr-Muller16}). 
The incident photon flux $\Phi(\mathrm{H})$ is defined as
\begin{equation}
    \Phi({\rm H}) = \frac{Q({\rm H})}{4\pi R^2}, 
\end{equation}
where $Q({\rm H})$ is the ionizing-photon production rate and $R$ is the distance between the ionizing source and the illuminated cloud face. 
The corresponding ionization parameter at distance $R$ is
\begin{equation}
    U = \frac{\Phi(\mathrm{H})}{n_{\rm H}c}, 
\end{equation}
where $c$ is the speed of light. 
The $\Phi(\mathrm{H})$--$n_{\rm H}$ plane is widely used to parameterize AGN broad-line-region (BLR) gas \citep[e.g.,][]{korista04, Schnorr-Muller16, kim18}.

The adopted model setup, parameter ranges, and grid spacing are summarized in Table~\ref{tab:cloudy}. At each point in the $\Phi(\mathrm{H})$--$n_{\rm H}$ plane, we calculate models over a range of total hydrogen column densities ($N_{\rm H}$), stopping each calculation at the specified $N_{\rm H}$. We adopt gas metallicities of $Z/Z_\odot=0.01, 0.1$, and $1.0$. The incident continuum is the AGN SED implemented in \textsc{Cloudy} \citep{mathews87}, with its normalization set by $\Phi(\mathrm{H})$. We include no microturbulence in the fiducial models. 

For each model, we calculate Balmer and Paschen line ratios directly from the line intensities reported by \textsc{Cloudy}, rather than measuring them from the synthetic model SEDs \citep[e.g.,][]{yan26}. 
This approach avoids additional dependencies on the assumed spectral resolution, line profile, continuum placement, and integration windows. 
The synthetic spectra around H$\beta$ contain numerous blended emission features, including iron lines and a structured pseudo-continuum. Measuring H$\beta$ directly from the model SEDs would introduce an unnecessary dependence on the adopted models. 
We compare the intrinsic model line ratios with the observed flux ratios measured in Section~\ref{subsec:flux-measurements}. We discuss the comparisons in Section~\ref{sec:discussion} and show the model grid results in Appendix~\ref{appendix:cloudy-model}. 

We note that there are other choices for geometries and line-output conventions.  For the fiducial calculations, we adopt plane-parallel geometry and obtain the line ratios using the \textsc{Cloudy} ``total" line intensity, $I_{\rm total} = I_{\rm inward} + I_{\rm outward}$. This quantity represents the total two-sided line emission from the slab, usually used in AGN studies, rather than the directional line flux seen by an observer from one side. We adopt this convention because it provides an approximate reproduction of the observed $L_{\rm bol}$--$L_{\rm broad,~H\alpha}$ scaling relation of LRDs \citep{yanagisawa26}. 

We also repeat representative calculations using the \textsc{Cloudy} \texttt{sphere} command, which approximates closed geometry. The qualitative dependence of the Balmer and Paschen ratios on gas density, ionizing photon flux, and column density does not change. Detailed comparisons among geometries and line-output conventions are presented in Appendix~\ref{appendix:cloudy-model-output}. We note that \citet{yan26} derive their predicted Balmer line ratios from the outward line-emission component. We also discuss them in Appendix~\ref{appendix:cloudy-model-output}. 

\begin{deluxetable*}{lcl}
    \tablecaption{Fiducial \textsc{Cloudy} models. \label{tab:cloudy}}
    \tabletypesize{\normalsize}
    \tablewidth{0pt}
    \tablehead{
    \colhead{Parameter} & \colhead{Adopted value or range} & \colhead{Description}
    } 
    \startdata
    Geometry & Plane-parallel slab & Constant-density gas geometry \\
    Incident continuum & AGN & Ionizing radiation field \\
    Metallicity & $0.01$, $0.1$, and $1.0~Z_\odot$ & Gas-phase metallicity \\
    $\log(n_\mathrm{H}/\mathrm{cm^{-3}})$ & 5--16, in steps of 0.5 dex & Hydrogen number density \\
    $\log(\Phi(\mathrm{H})/\mathrm{cm^{-2}~s^{-1}})$ & 16--24, in steps of 0.5 dex & Incident ionizing photon flux \\
    $\log(N_\mathrm{H}/\mathrm{cm^{-2}})$ & 21--26, in steps of 0.5 dex & Stopping hydrogen column density \\
    Dust grains & None & Intrinsic, dust-free line-ratio calculation \\
    Microturbulence & None & Turbulence \\
    \enddata
    \tablecomments{The incident AGN continuum is based on \citet{mathews87}. Although the full calculations extend to $\log(n_{\rm H}/{\rm cm^{-3}})=16$, Figures~\ref{fig:HaHb_PabPag_cloudy} and \ref{fig:PabPag_PaaPab_cloudy} display only $\log(n_{\rm H}/{\rm cm^{-3}})\leq12$ for clarity.}
\end{deluxetable*}

\section{Results} \label{sec:results} 

\subsection{Broad and Narrow Balmer Decrements} 

%%% fig: broadHaHb vs. Haluminosity %%%
\begin{figure}
% \plotone{figures/fig_broadHaHb_Haluminosity_v5_with_lin2026b_logy.pdf}
\includegraphics[width=1.0\linewidth]{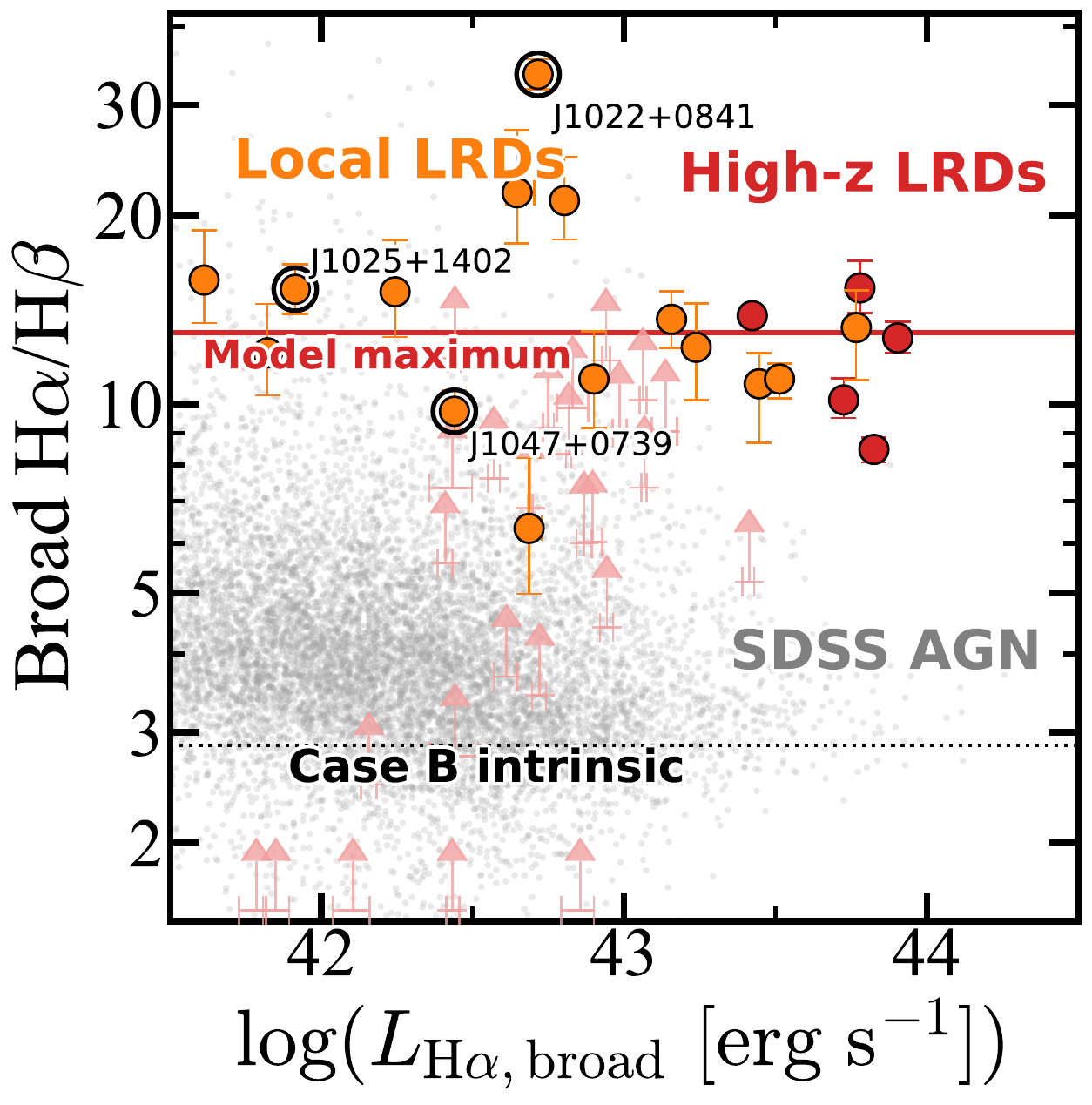}
    \caption{
    Relation between the broad H$\alpha$/H$\beta$ ratio and the broad H$\alpha$ luminosity. The red and orange circles show high-redshift and low-redshift LRDs with both H$\alpha$ and H$\beta$ ${\rm S/N}\geq5$ listed in Table~\ref{tab:sample}, respectively. The red arrows show the high-redshift LRDs without $\geq5\sigma$ broad H$\beta$ detections. The gray points show the SDSS AGNs at $z<0.35$ \citep{liu19}. The Case B intrinsic value ($=2.87$) is shown as a black dotted line. The red line indicates the maximum ratio ($\sim13$) obtained in the dust-free \textsc{Cloudy} grid (Section~\ref{sec:discussion}). The three low-redshift LRDs with available Paschen-line and MIR data are highlighted with black circles. 
    }
\label{fig:broadHaHb_Haluminosity}
\end{figure}
%%%%%%%%%%%%%%%%%%%%%%%%%%%%%%%%%%%%%%%

We combine the high-redshift line measurements from Section~\ref{subsec:flux-measurements} with the published low-redshift fluxes described in Section~\ref{subsec:local-sample}. 
We first examine the broad Balmer decrements, which are widely observed and linked to the central engine of LRDs. 

Figure~\ref{fig:broadHaHb_Haluminosity} shows the relation between the broad H$\alpha$/H$\beta$ ratio and the broad H$\alpha$ luminosity ($L_{\rm H\alpha,\,broad}$). 
The red and orange circles show high- and low-redshift LRDs, respectively, with both broad H$\alpha$ and broad H$\beta$ detected at ${\rm S/N}\geq5$. The red arrows indicate lower limits on the broad H$\alpha$/H$\beta$ ratio for high-redshift LRDs with broad H$\beta$ to have ${\rm S/N}<5$. The gray points show a comparison sample of SDSS AGNs ($z<0.35$; \citealt{liu19}). 
For reference, unobscured Type-1 AGNs typically have intrinsic broad H$\alpha$/H$\beta$ ratios of $\sim2.7$--$3.1$, close to the Case B value \citep{dong08, gaskell17}. 

In Figure~\ref{fig:broadHaHb_Haluminosity}, most LRDs have broad H$\alpha$/H$\beta$ ratios above the intrinsic Case B recombination value ($=2.87$ for $n_e=10^2~{\rm cm^{-3}}$ and $T_e=10^4$~K; e.g., \citealt{storey95}), and several reach H$\alpha$/H$\beta\gtrsim13$. These extreme values exceed the maximum ratio ($\mathrm{H\alpha/H\beta}\sim13$) in our adopted dust-free \textsc{Cloudy} grid shown as the red line in Figure~\ref{fig:broadHaHb_Haluminosity}. The red line at H$\alpha$/H$\beta\sim13$ marks the largest ratio reached in the parameter space explored by the dust-free Cloudy models (open plane-parallel slabs and total line emission from both faces) discussed in Section~\ref{sec:discussion} and shown in Figures~\ref{fig:lognH_HaHb} and \ref{fig:logNH_HaHb}. The implications of this comparison, including the possible role of attenuation, are discussed in Section~\ref{sec:discussion}. LRDs also tend to have higher broad H$\alpha$/H$\beta$ values than the SDSS comparison sample (gray; \citealt{liu19}, see also \citealt[]{geris26}). Some SDSS AGNs also show H$\alpha$/H$\beta>13$. A detailed analysis of these SDSS AGNs with high H$\alpha$/H$\beta$ is beyond the scope of this study and will be presented in a forthcoming paper. We briefly discuss them in Appendix~\ref{appendix:sdss_high_HaHb}, together with comparisons to LRDs. 

%%% fig: broadHaHb vs. narrowHaHb %%%
\begin{figure}
    \includegraphics[width=1.0\linewidth]{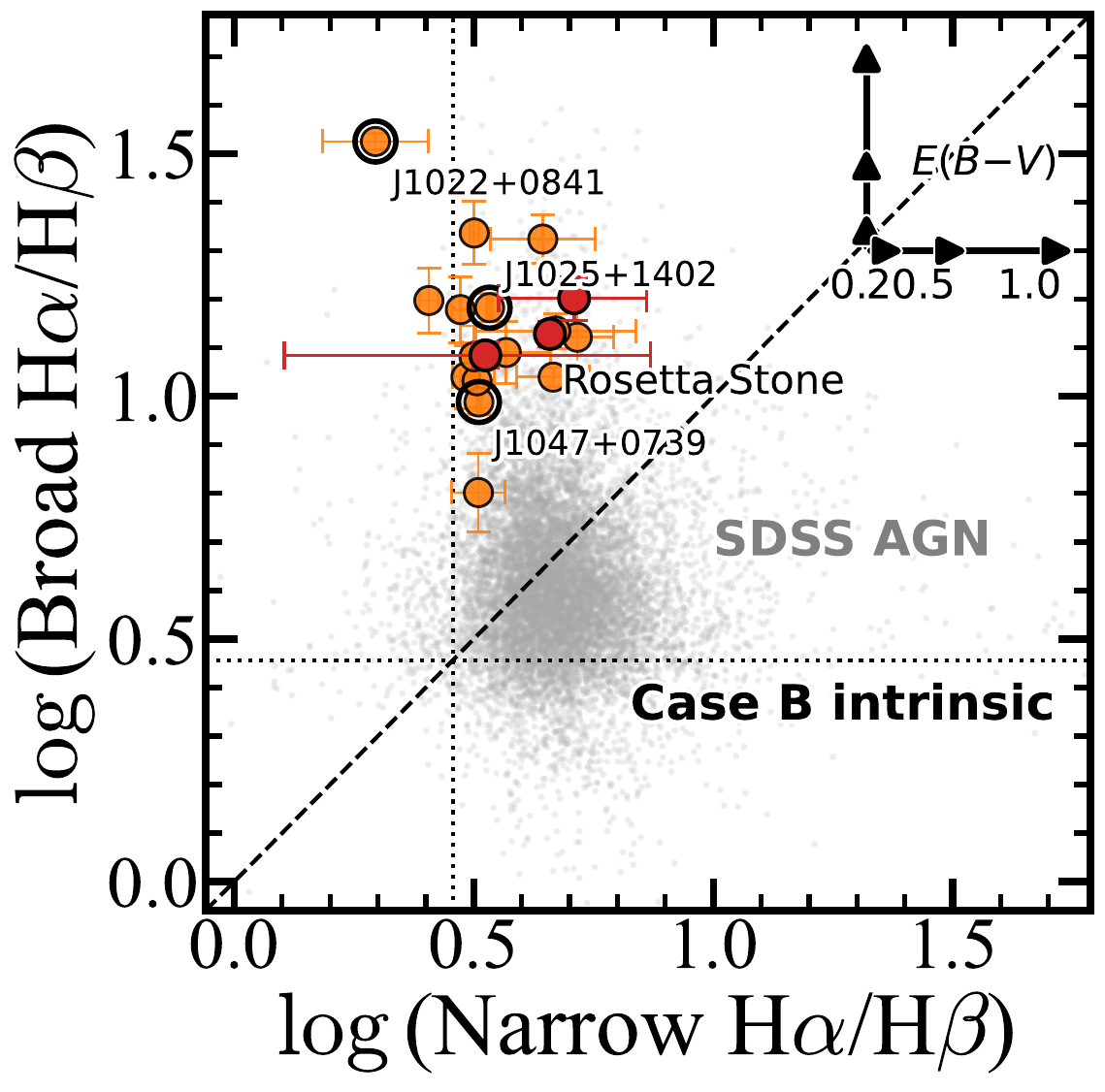}
    \caption{
    Relation between the broad H$\alpha$/H$\beta$ ratio and the narrow H$\alpha$/H$\beta$ ratio. The red and orange circles show high-redshift and low-redshift LRDs with both H$\alpha$ and H$\beta$ ${\rm S/N}\geq5$, respectively. The arrows show the $E(B-V)$ values (0.2, 0.5, and 1.0) assuming the Small Magellanic Cloud (SMC) extinction curve. The black dashed line shows equality between the broad and narrow ratios. The Case B intrinsic value ($=2.87$) is shown as the black dotted line. The other symbols are the same as in Figure~\ref{fig:broadHaHb_Haluminosity}. The Rosetta Stone (JADES-GN-28074) is also labeled.
    }
\label{fig:broadHaHb_narrowHaHb}
\end{figure} 
%%%%%%%%%%%%%%%%%%%%%%%%%%%%%%%%%%%%%%

Comparison between the broad line ratio and the narrow one provides insight into the physical properties of LRDs. 
Figure~\ref{fig:broadHaHb_narrowHaHb} compares the broad and narrow H$\alpha$/H$\beta$ ratios. 
The main symbols are the same as in Figure~\ref{fig:broadHaHb_Haluminosity}.
For reference, we also show the corresponding $E(B-V)$ values based on the H$\alpha$/H$\beta$ ratio as black arrows, assuming the Small Magellanic Cloud (SMC) attenuation curve \citep{gordon03} implemented in the \texttt{dust\_extinction} package \citep{Gordon2024}. 
The different extinction curves (e.g., \citealt{calzetti00}, the Large Magellanic Cloud, and the Milky Way; see also \citealt{gaskell07}) show different $E(B-V)$ values for the Balmer and Paschen decrements, but changing the extinction curve affects the inferred values of $E(B-V)$ but not the qualitative trends discussed below. In the following of this paper, we use the SMC law \citep{gordon03} as a fiducial extinction curve. 

In Figure~\ref{fig:broadHaHb_narrowHaHb}, for both high- and low-redshift samples, the broad-line ratios are generally larger than the corresponding narrow-line ratios \citep[see also][]{lin26, nikopoulos25, brooks25}. The narrow H$\alpha$/H$\beta$ ratios almost follow the Case B intrinsic value ($=2.87$). 
This trend demonstrates that the broad-line emitting region has some non-Case B effects and/or dust extinction, while the narrow-line emitting region generally follows no/little dust conditions.

\subsection{Broad Balmer and Paschen Line Ratios}  

%%% fig: broad Ha/Hb vs. broad Pab/Pag %%%
\begin{figure*}
    % \plotone{figures/fig_HaHb_PabPag_v3.pdf}
    \plotone{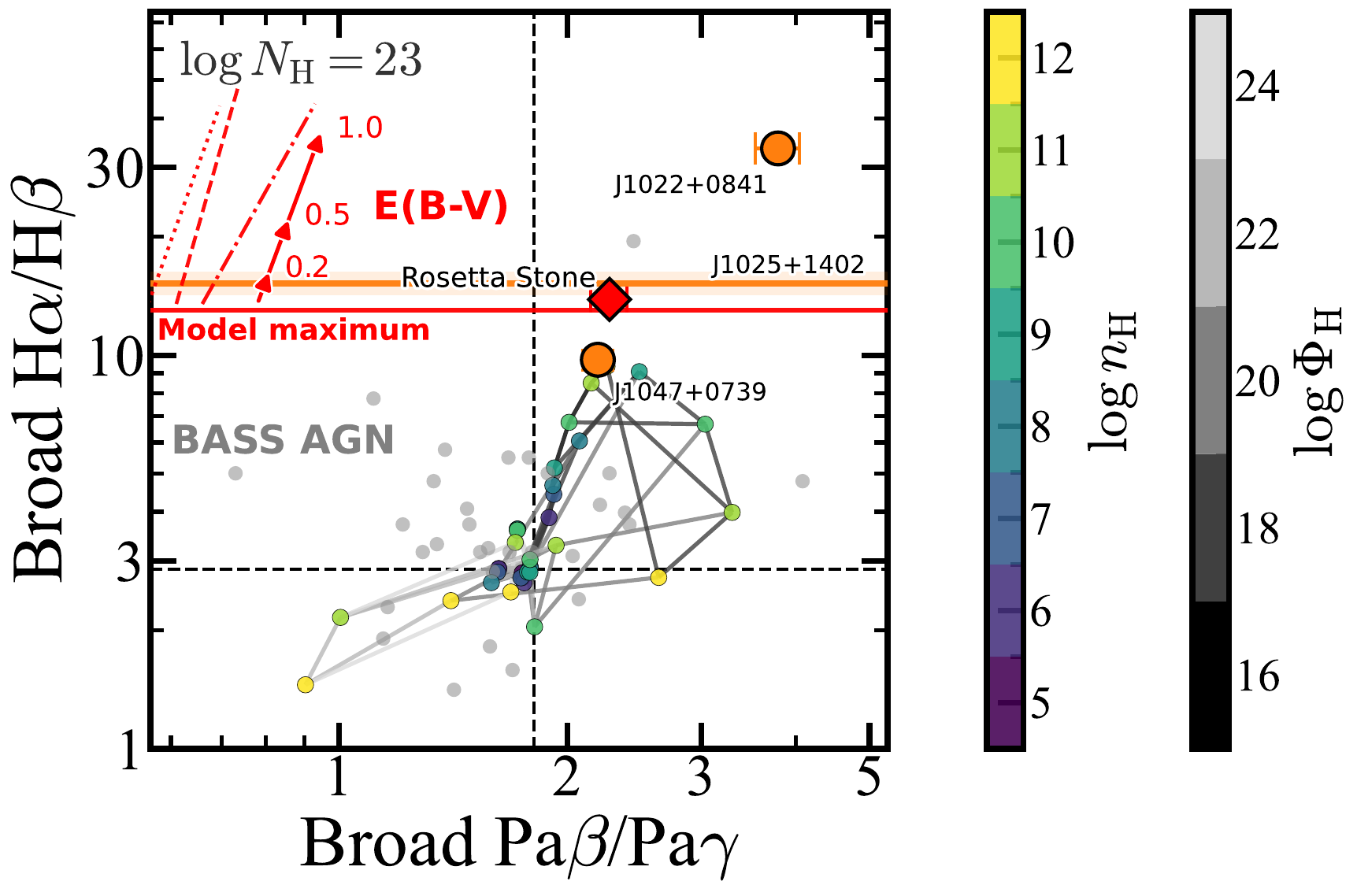}
    \caption{
    Relation between the broad H$\alpha$/H$\beta$ ratio and the broad Pa$\beta$/Pa$\gamma$ ratio. The red diamond and orange circles show the high-redshift and low-redshift LRDs, respectively. 
    The orange line shows the H$\alpha$/H$\beta$ of J1025+1402, which has no reliable Pa$\beta$ flux.
    The gray circles show local AGNs from the Swift Burst Alert Telescope (BAT) AGN Spectroscopic Survey (BASS; \citealt{lamperti17, denbrok22, mejia-Restrepo22}). 
    The colored points show the dust-free \textsc{Cloudy} models with $\log(n_{\rm H}/{\rm cm}^{-3})=5$--$12$, as indicated by the color bar. 
    The gray lines connect models with $\log(\Phi({\rm H})/{\rm cm}^{-2}~{\rm s}^{-1})=16$, 18, 20, 22, and 24. 
    The models assume $\log(N_{\rm H}/{\rm cm}^{-2})=23$ and $Z/Z_\odot=0.1$. 
    The red line shows the maximum H$\alpha$/H$\beta$ values in our model framework. 
    The red arrow indicates the reddening vector for the SMC attenuation curve ($E(B-V)=0.2, 0.5$, and $1.0$). 
    The dotted, dashed, and dash-dotted lines, from left to right in the top left of this figure, show $E(B-V)=1.0$ in the case of the Milky Way, the Large Magellanic Cloud (LMC), and the \citet{calzetti00} law, respectively. 
    The horizontal and vertical dashed lines show the Case~B intrinsic value for H$\alpha$/H$\beta$ and Pa$\beta$/Pa$\gamma$, respectively. 
    }
\label{fig:HaHb_PabPag_cloudy}
\end{figure*}
%%%%%%%%%%%%%%%%%%%%%%%%%%%%%%%%%%%%%%%%%%

A subset of the LRD spectra covers broad Paschen emission lines. In this subsection, we summarize the broad Balmer and Paschen line ratios. Figure~\ref{fig:HaHb_PabPag_cloudy} compares broad H$\alpha$/H$\beta$ with broad Pa$\beta$/Pa$\gamma$. The usable measurements comprise two low-redshift LRDs (J1022+0841 and J1047+0739) and one high-redshift LRD, the Rosetta Stone \citep{juodvbalis24}. Although Pa$\beta$ is covered for the local LRD J1025+1402, strong telluric absorption prevents a reliable measurement \citep{lin26}. We therefore exclude this source from the Pa$\beta$/Pa$\gamma$ comparison. 

The LRDs have broad Pa$\beta$/Pa$\gamma$ ratios spanning $\sim1.5$--$4$. In the H$\alpha$/H$\beta$--Pa$\beta$/Pa$\gamma$ plane, the position of J1047+0739 is consistent with the direction expected for a simple SMC foreground screen, assuming the adopted intrinsic ratios. The remaining LRDs do not lie along this simple reddening locus. Because broad-line intrinsic ratios can depart from Case B, this comparison alone does not uniquely determine the attenuation. 
For comparison, we also show the local AGN sample selected by the hard X-ray Swift-Burst Alert Telescope (BAT) AGN Spectroscopic Survey \citep[BASS;][]{lamperti17, denbrok22, mejia-Restrepo22} as gray circles. 
A subset of the BASS AGNs overlaps with the LRDs in the broad Balmer and Paschen line-ratio diagram. This overlap suggests that some local AGNs may share similar line-forming conditions, such as high density, large line optical depths, and/or differential attenuation \citep[see also][]{ricci22}. 

%%% fig: broad Pab/Pag vs. broad Paa/Pab %%%
\begin{figure*}
    \plotone{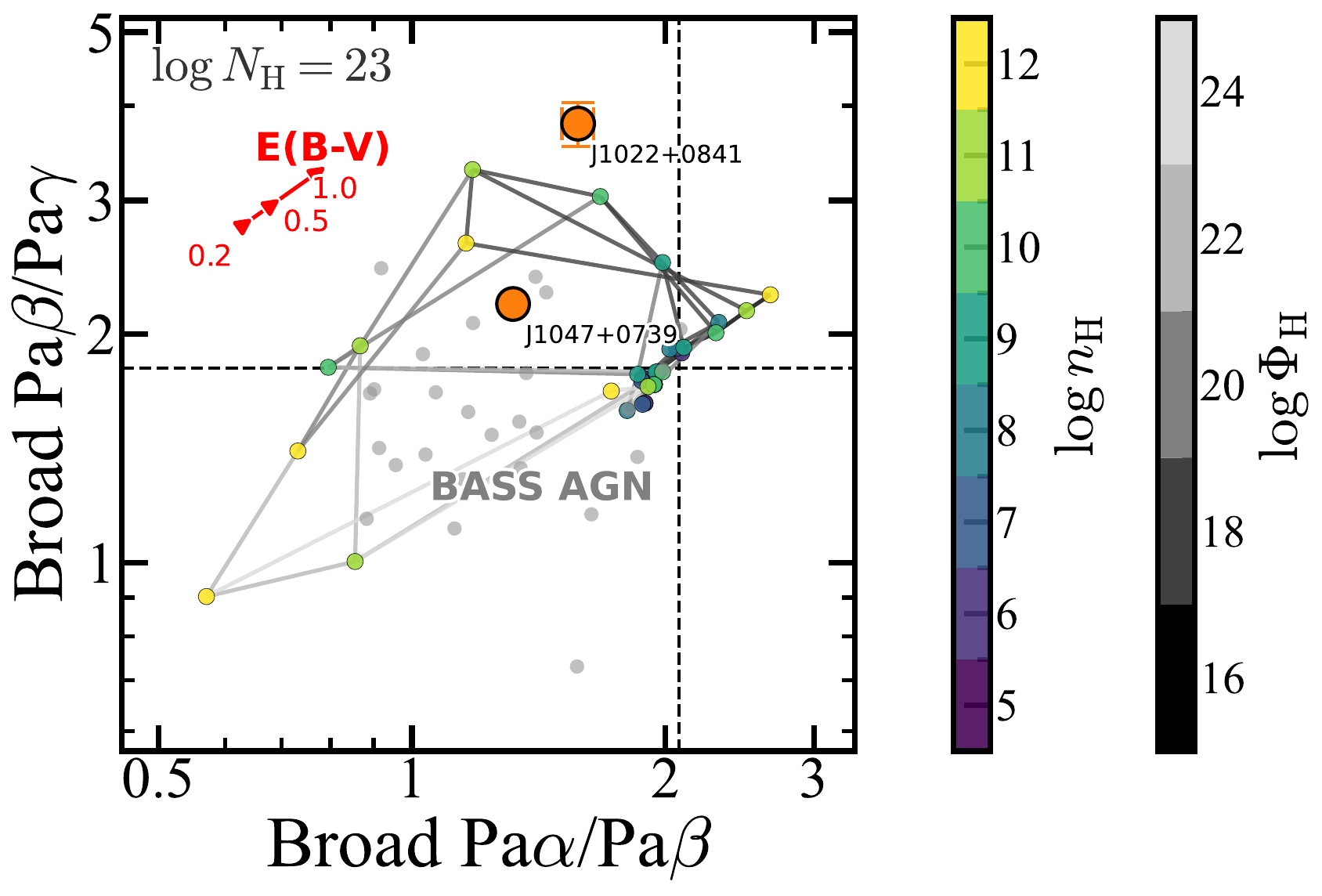}
    \caption{
    Relation between the broad Pa$\beta$/Pa$\gamma$ ratio and the broad Pa$\alpha$/Pa$\beta$. The symbols and the dust-free \textsc{Cloudy} model grids are the same as in Figure~\ref{fig:HaHb_PabPag_cloudy}. 
    }
\label{fig:PabPag_PaaPab_cloudy}
\end{figure*}
%%%%%%%%%%%%%%%%%%%%%%%%%%%%%%%%%%%%%%

Figure~\ref{fig:PabPag_PaaPab_cloudy} compares broad Pa$\beta$/Pa$\gamma$ with broad Pa$\alpha$/Pa$\beta$. Only J1022+0841 and J1047+0739 are shown because they are the only LRDs in our sample with reliable measurements of all three broad Paschen lines. The symbols and model grids have the same meanings as in Figure~\ref{fig:HaHb_PabPag_cloudy}. Both sources have broad Pa$\alpha$/Pa$\beta$ ratios below the adopted Case B value. A standard foreground attenuation screen would instead increase this ratio, because Pa$\beta$ is more strongly attenuated than Pa$\alpha$. The observed ratios therefore cannot be explained by foreground attenuation alone under the Case B assumption. 
We note that RUBIES-UDS-40579, which is not shown in Figures~\ref{fig:HaHb_PabPag_cloudy} and \ref{fig:PabPag_PaaPab_cloudy} because it lacks broad H$\beta$ coverage, has Pa$\beta$/Pa$\gamma=1.46$. If it also has a large H$\alpha$/H$\beta$ ratio, the two ratios may challenge the models or indicate dust extinction, motivating future H$\beta$ spectroscopy. 
The implications of the \textsc{Cloudy} model comparison in Figure~\ref{fig:HaHb_PabPag_cloudy} and Figure~\ref{fig:PabPag_PaaPab_cloudy} are discussed in Section~\ref{sec:discussion}.

\section{Discussion} \label{sec:discussion}

In this section, we interpret the observed broad hydrogen-line ratios using the \textsc{Cloudy} calculations described in Section~\ref{subsec:cloudy}. We first assess whether optically thick, high-density line-forming gas can reproduce the observed broad-line ratios without attenuation. We then examine whether additional line-of-sight attenuation is required within the assumptions of our model framework. The implications for the dust distribution are discussed together with the MIR comparisons in the following subsections.

\subsection{Comparison with \textsc{Cloudy} Models} \label{subsec:cloudy-comparison}

Figures~\ref{fig:HaHb_PabPag_cloudy} and \ref{fig:PabPag_PaaPab_cloudy} show representative \textsc{Cloudy} model grids with $Z/Z_\odot=0.1$ and $\log(N_{\rm H}/{\rm cm}^{-2})=23$. The colored points span $\log(n_{\rm H}/{\rm cm}^{-3})=5$--$12$, and the gray curves show models with $\log(\Phi({\rm H})/{\rm cm}^{-2}~{\rm s}^{-1})=16$--$24$. 

In Figure~\ref{fig:HaHb_PabPag_cloudy}, the models predict elevated H$\alpha$/H$\beta$ ratios at $\log(n_{\rm H}/{\rm cm}^{-3})\sim10$ and $\log(\Phi({\rm H})/{\rm cm}^{-2}~{\rm s}^{-1})\sim18$, where H$\alpha$/H$\beta$ can reach $\sim10$ (see also Appendix~\ref{appendix:cloudy-model} for the full model grids). 

In this regime, the hydrogen-line ratios are modified by the combined effects of large line optical depths, radiative trapping, and collisional processes, which alter the level populations and emergent line emissivities relative to the Case B approximation \citep[e.g.,][]{wills85, ilic12, ruff12, wu23, son25, yan26, chang26}. While Case B assumes efficient trapping of Lyman-series photons but optically thin Balmer and Paschen transitions, the present models can also become optically thick in the Balmer lines. In particular, repeated absorption and re-emission of trapped H$\beta$ photons can alter the branching of the $n=4$ level. A photon initially emitted as H$\beta$ ($4\rightarrow2$) may instead follow the $4\rightarrow3\rightarrow2$ cascade, producing Pa$\alpha$ and H$\alpha$. In addition, collisional excitation ($2\rightarrow3$) and de-excitation at high density may further contribute to the departure from the Case-B ratios. \citet{kwan81} show that collisional processes and line trapping in optically thick, photoionized clouds modify hydrogen-level populations and intrinsic Balmer and Paschen line ratios. More recently, \citet{chang26} explicitly model the H$\beta$ conversion described above, demonstrating that repeated scattering can suppress H$\beta$ while producing Pa$\alpha$ and H$\alpha$, thereby increasing H$\alpha$/H$\beta$ without dust extinction. The magnitude of this enhancement depends on the line optical depth and the intrinsic width of the incident emission.

The Pa$\alpha$/Pa$\beta$ ratio (Figure~\ref{fig:PabPag_PaaPab_cloudy}) provides a complementary test of intrinsic departures from Case~B, which cannot be uniquely distinguished from dust attenuation using H$\alpha$/H$\beta$ alone. A foreground attenuation screen increases Pa$\alpha$/Pa$\beta$, because Pa$\beta$ is more strongly attenuated than Pa$\alpha$ (red arrows in Figure~\ref{fig:PabPag_PaaPab_cloudy}). In contrast, in optically thick, high-density gas, the Pa$\alpha$ ($4\rightarrow3$) transition can become optically thick before Pa$\beta$ ($5\rightarrow3$). The escape probability of Pa$\alpha$ is then reduced more strongly by photon trapping and radiative or collisional redistribution, allowing the emergent Pa$\alpha$/Pa$\beta$ ratio to fall below its Case-B value. Therefore, a ratio below the adopted Case B value cannot be produced by foreground attenuation alone if the intrinsic ratio is assumed to be Case B. Among the sources with sufficient Paschen-line measurements, both J1022+0841 and J1047+0739 have broad Pa$\alpha$/Pa$\beta$ ratios below this value (Figure~\ref{fig:PabPag_PaaPab_cloudy}). These measurements support non-Case-B hydrogen-line formation in their broad-line-emitting gas and are consistent with the optically thick, high-density regime (e.g., \citealt{wills85, drake80, glikman06, landt08, soifer04}). 

%% fig: Ha/Hb vs. lognH %%%%%%%%%%%%%%%%%%
\begin{figure*}
    \plotone{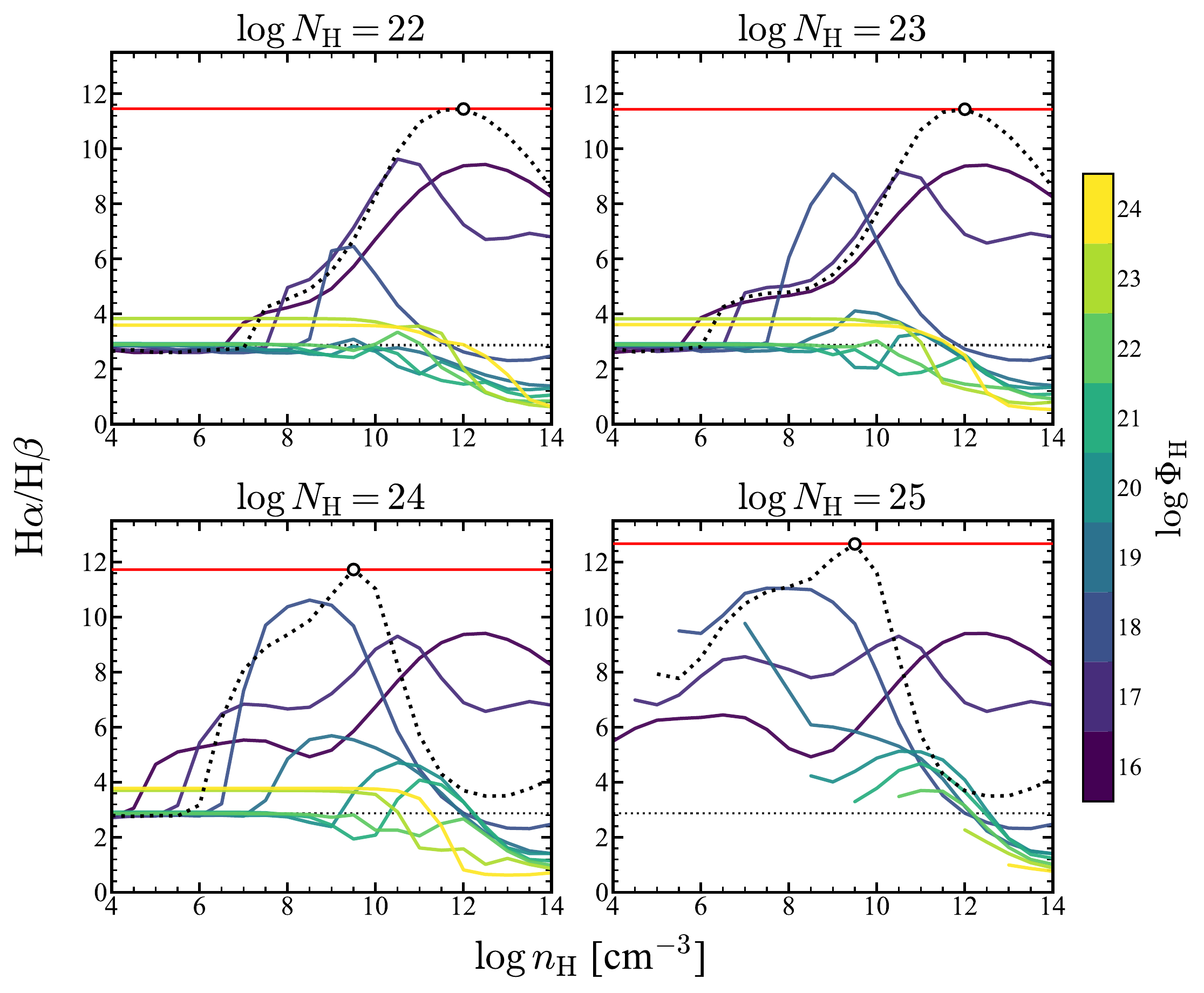}
    \caption{
    Relation between the H$\alpha$/H$\beta$ ratio and the hydrogen number density ($\log n_\mathrm{H}$) obtained from the \textsc{Cloudy} modeling. The different lines show the different hydrogen column densities ($\log(N_{\rm H}/{\rm cm^{-2}})=22$--$25$). In this figure, we fix the ionizing photon flux $\Phi({\rm H})$ indicated by the color bar. The horizontal black dotted line shows the Case B intrinsic value. The black dashed curve shows the case for the highest H$\alpha$/H$\beta$ ratio in each column density at a fixed $\Phi({\rm H})$. The red lines show the maximum H$\alpha$/H$\beta$ in each column density.
    }
\label{fig:lognH_HaHb}
\end{figure*}
%%%%%%%%%%%%%%%%%%%%%%%%%%%%%%%%%%%%%%

%%% fig: logNH vs. Ha/Hb %%%%%%%%%%%%%
\begin{figure}
    \includegraphics[width=1.0\linewidth]{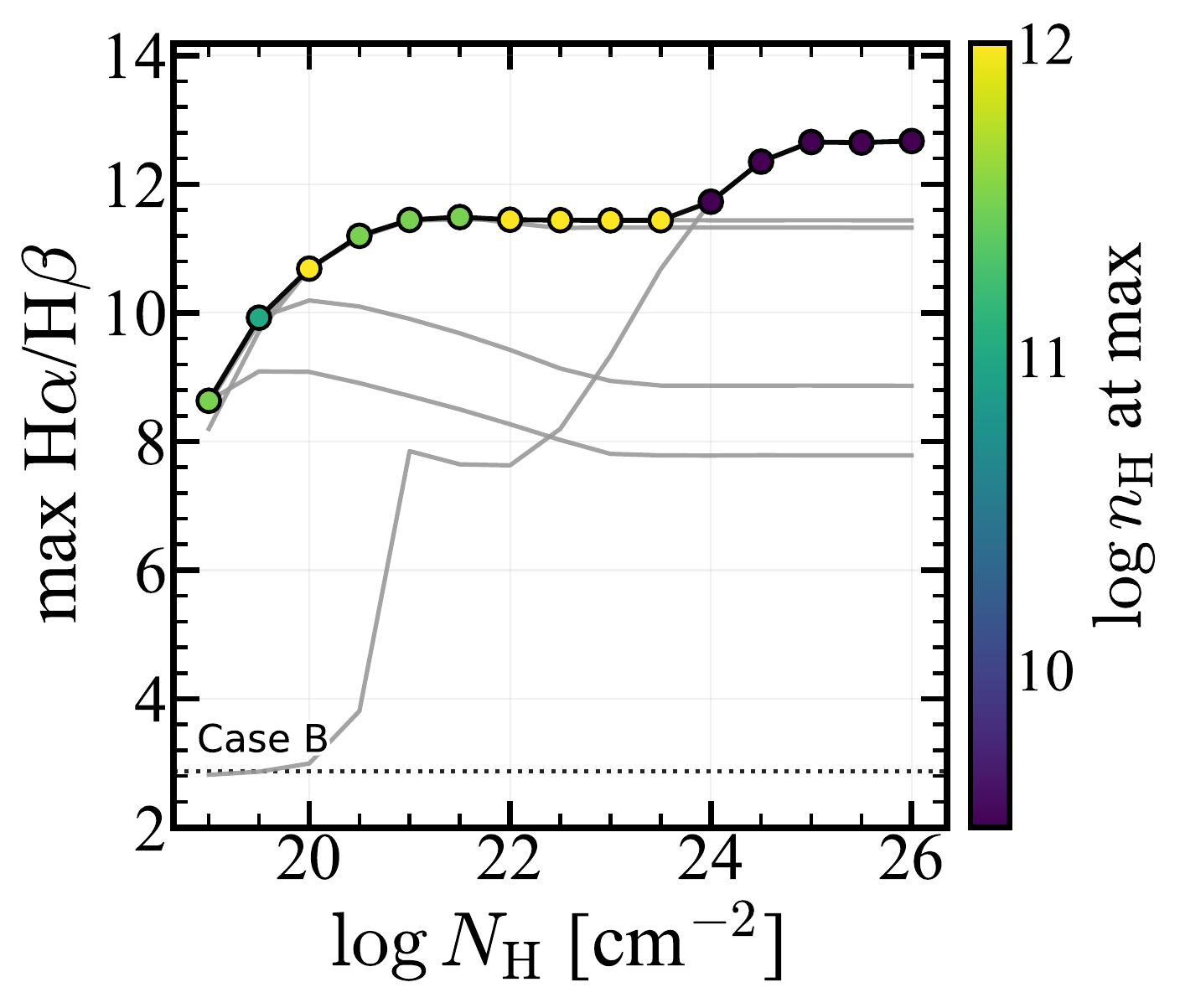}
    % \plotone{figures/fig_HaHb_logNH_cloudy_max_v1.pdf}
    \caption{ 
    Relation between the H$\alpha$/H$\beta$ ratio and the hydrogen column density ($\log N_{\rm H}$) obtained from the \textsc{Cloudy} models. At each column density, we plot the maximum H$\alpha$/H$\beta$ value obtained among the model grids. The color of each data point indicates the hydrogen number density that yields this maximum ratio, as shown by the colorbar. The gray curves show the H$\alpha$/H$\beta$ ratios as a function of $\log N_{\rm H}$ at fixed $(\log n_{\rm H},\log\Phi{\rm (H)})$ for the models that reach the maximum at one or more column densities. The horizontal black dotted line denotes the Case B recombination value. 
    }
\label{fig:logNH_HaHb}
\end{figure} 
%%%%%%%%%%%%%%%%%%%%%%%%%%%%%%%%%%%%%%

We also note that these line ratios remain largely unchanged even when we choose other high column densities ($\log(N_{\rm H}/{\rm cm^{-2}})=22$--$25$) and metallicities ($Z/Z_\odot=0.01$ and $1.0$). The column-density dependence becomes weak once the cloud is ionization-bounded. Beyond the hydrogen ionization front, additional deeper gas contributes little to the main Balmer-line-emitting region, causing the line ratios to saturate with increasing $N_{\rm H}$. The Balmer decrements primarily depend on hydrogen level populations and line optical depths and show only a weak net dependence on metallicity. Figure~\ref{fig:lognH_HaHb} shows the relation between the number density $\log n_{\rm H}$ and the H$\alpha$/H$\beta$ ratio obtained from the models. At fixed $\Phi(\mathrm{H})$, the ratio increases toward the intermediate densities, reaches a maximum, and then declines at higher densities. The turnover occurs as Pa$\alpha$ also becomes optically thick, reducing the efficiency of the H$\beta\rightarrow{\rm Pa}\alpha+{\rm H}\alpha$ conversion. At still higher densities, collisional excitation and de-excitation drive the excited-state populations toward LTE, and the H$\alpha$/H$\beta$ ratio is lower than the maximum value. The maximum values of H$\alpha$/H$\beta$ ratios in our grid are $\sim13$, and even if we choose a different column density, the general trend of the maximum H$\alpha$/H$\beta$ does not change. 

To further test the effects of column density, we also show the relation between $\log N_{\rm H}$ and the H$\alpha$/H$\beta$ ratio in Figure~\ref{fig:logNH_HaHb}. Figure~\ref{fig:logNH_HaHb} shows the maximum H$\alpha$/H$\beta$ ratios for each column density, indicating that the H$\alpha$/H$\beta$ ratio is almost saturated for high column density. Thus, we find that H$\alpha$/H$\beta\sim13$ is the threshold achievable without dust, accounting for non-Case B effects in the \textsc{Cloudy} model framework. We show the resulting full grid H$\alpha$/H$\beta$ ratio of the $\Phi({\rm H})$--$n_{\rm H}$ plane in Appendix~\ref{appendix:cloudy-model}. We note that at higher column density ($\log(N_\mathrm{H}/\mathrm{cm^{-2}})>24$), \textsc{Cloudy} calculations sometimes become unstable, but the general trend does not change and the main discussion of this paper remains unchanged. 

Building on these results, we examine and summarize how the observed line ratios of individual sources can be explained. The H$\alpha$/H$\beta$, Pa$\alpha$/Pa$\beta$, and Pa$\beta$/Pa$\gamma$ ratios of J1022+0841 are simultaneously explained by the non-Case B model of $\log(n_\mathrm{H}/\mathrm{cm^{-3}})\sim10$--$11$, $\log(\Phi_\mathrm{H}/\mathrm{cm^{-2}~s^{-1}})\sim18$ and $E(B-V)\sim1$ in Figures~\ref{fig:HaHb_PabPag_cloudy} and \ref{fig:PabPag_PaaPab_cloudy}. These line ratios of another local LRD, J1047+0739, are also explained by the similar parameter space, but without additional attenuation required (Figures~\ref{fig:HaHb_PabPag_cloudy} and \ref{fig:PabPag_PaaPab_cloudy}). For the Rosetta Stone, the H$\alpha$/H$\beta$--Pa$\beta$/Pa$\gamma$ comparison identifies an approximately compatible model region near $\log(n_\mathrm{H}/\mathrm{cm^{-3}})\sim9$, $\log(\Phi_\mathrm{H}/\mathrm{cm^{-2}~s^{-1}})\sim18$, and $\log N_{\rm H}=25$ model grids with $E(B-V)\sim0.2$. These results also highlight that non-Case B physics alters Balmer and Paschen line ratios, but some objects still require additional attenuation toward the broad-line-emitting gas to fully explain the Balmer and Paschen line ratios simultaneously. 
The joint Balmer-Paschen comparison allows J1047+0739 to be explained without additional attenuation, while J1022+0841 and the Rosetta Stone require it within the models. 

For sources without usable Paschen constraints, we apply the maximum H$\alpha$/H$\beta\sim13$ as a diagnostic of the additional attenuation.
Two high-redshift and eight low-redshift sources exceed this threshold (red line in Figure~\ref{fig:broadHaHb_Haluminosity}), corresponding to ten of the twenty primary-sample objects (Figure~\ref{fig:broadHaHb_Haluminosity}).
Except for J1022+0841 and the Rosetta Stone, these sources lack usable joint Balmer-Paschen constraints.
Within the adopted geometry and line-output prescription, their extreme decrements favor additional attenuation to reach the H$\alpha$/H$\beta$ values above the dust-free model. 
For Balmer-only sources below the threshold, the relative contributions of non-Case-B effects and attenuation remain unclear.
Thus, the full sample establishes the occurrence of extreme Balmer decrements in the selected objects, while the three objects (J1047+0739, J1022+0841 and the Rosetta Stone), which have sufficient Paschen line measurements, provide stronger physical constraints. 

We also mention that the dense, optically thick gas invoked to explain the non-Case-B hydrogen-line ratios may also be related to the gas responsible for the frequent Balmer absorption (e.g., \citealt{chen26, davis26, juodvzbalis26, matthee26, yanagisawa26_atlas2}). 
Balmer absorption requires an appreciable line-of-sight column of neutral hydrogen in the $n=2$ state.
Such excited hydrogen also provides the Balmer opacity relevant to the radiative trapping discussed above, suggesting that the absorbing and line-emitting gas could belong to a common circumnuclear gas system. 
Alternatively, the absorption could arise in a physically distinct foreground component.
Our models describe the emitted hydrogen-line ratios but do not jointly constrain the absorption profiles and gas geometry. They therefore cannot establish whether the absorbing and emitting regions are co-spatial.

\subsection{Comparison with Mid-infrared Constraints} \label{sec:mid-infrared-constraints}

%%% fig: dust temperature %%%%%%%%%%%%%
\begin{figure*}
    \gridline{
        \fig{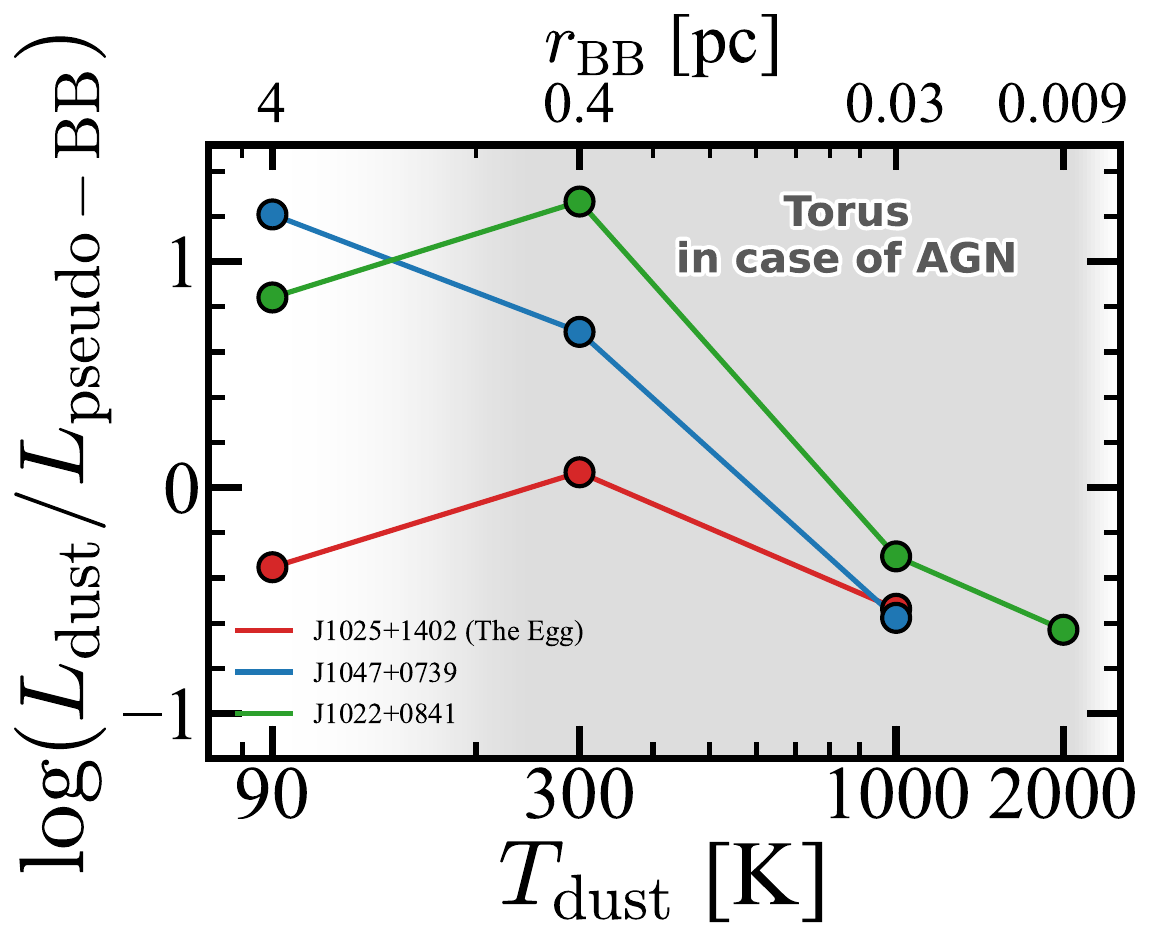}{0.54\textwidth}{}
        \fig{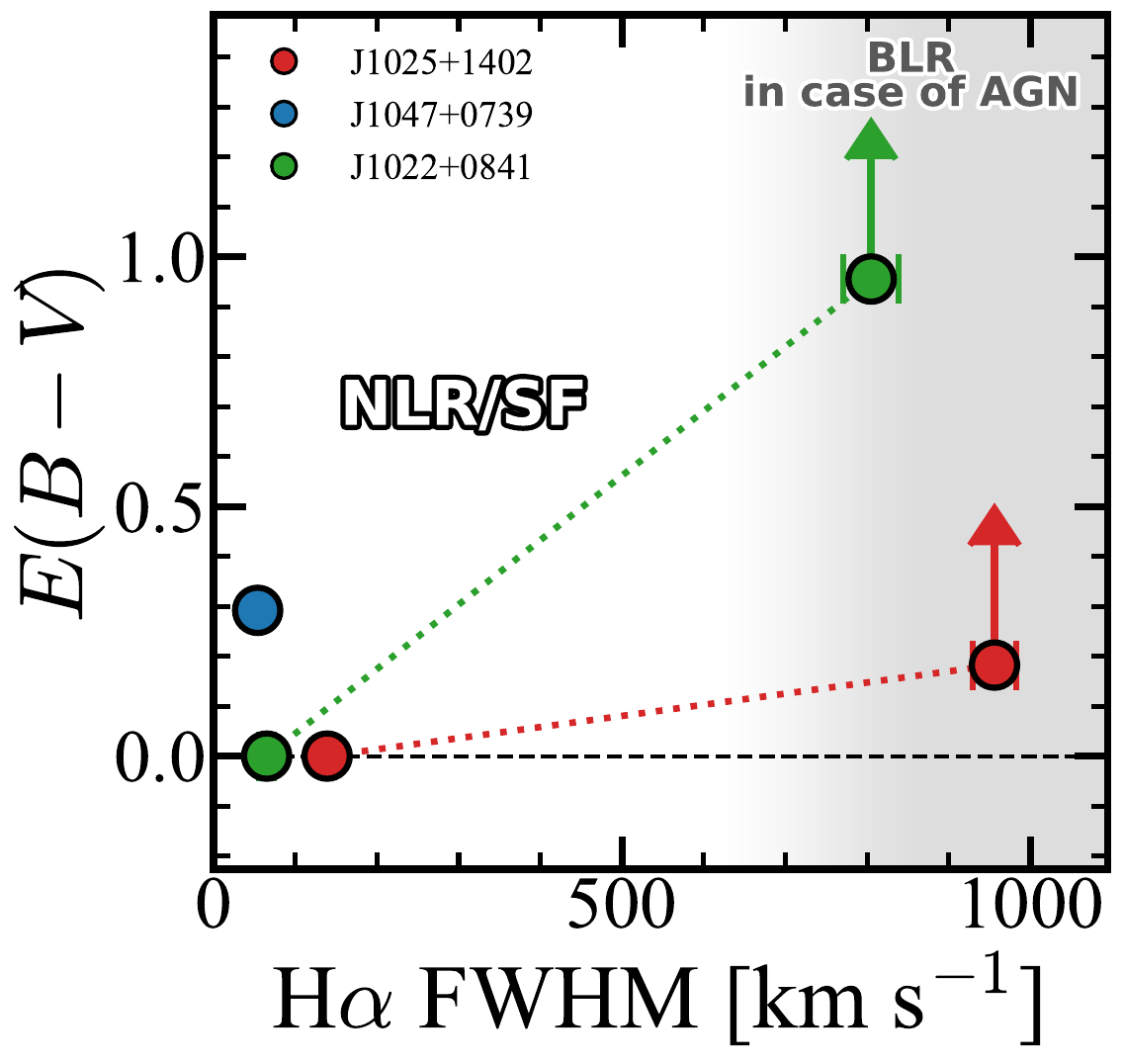}{0.41\textwidth}{}
    }
    \caption{
    Left: Relation between the dust luminosity normalized by the pseudo-blackbody luminosity, $L_{\rm dust}/L_{\rm pseudo-BB}$, and the dust-template temperature for the local LRDs studied by \citet{lin26}. 
    $L_{\rm dust}$ denotes the luminosity assigned to each fixed-temperature dust template in the SED decomposition of \citet{lin26}.
    The red, blue, and green symbols show J1025+1402, J1047+0739, and J1022+0841, respectively. The upper axis shows the corresponding blackbody-equilibrium heating radius ($r_\mathrm{BB}$) estimated from the $L_{\rm pseudo-BB}$ as an illustrative scale. 
    The shaded regions indicate the approximate temperature ranges commonly associated with circumnuclear tori and extended polar or narrow-line-region dust in AGNs \citep[e.g.,][]{lyu21}. 
    Right: Relation between the $E(B-V)$ and the H$\alpha$ FWHM. 
    The narrow-line $E(B-V)$ values are taken from \citet{lin26}, which are estimated from multiple narrow Balmer lines. 
    The broad-line $E(B-V)$ values show the minimum additional SMC-like attenuation required when the observed H$\alpha$/H$\beta$ ratio exceeds the maximum value in our dust-free \textsc{Cloudy} grid.
    The narrow-line FWHM values of J1025+1402 and J1047+0739 are slightly shifted for visualization. 
    J1047+0739 (blue) falls within the \textsc{Cloudy} model grids (Figures~\ref{fig:HaHb_PabPag_cloudy} and \ref{fig:PabPag_PaaPab_cloudy}), and we have no constraints on the extinction of broad lines. 
    }
\label{fig:dust_temperature}
\end{figure*}
%%%%%%%%%%%%%%%%%%%%%%%%%%%%%%%%%%%%%%

Broad H$\alpha$/H$\beta$ ratios that exceed the dust-free \textsc{Cloudy} grid can require additional differential attenuation within the model framework. 
As for local LRDs, Paschen line ratios (Figure~\ref{fig:PabPag_PaaPab_cloudy}) also suggest additional extinction.
MIR emission provides complementary evidence for dust around the central engine, but does not by itself identify the material responsible for broad-line attenuation.

\citet{lin26} fit the WISE SEDs of the three local LRDs with the fixed-temperature dust templates of \citet{lyu21}. The left panel of Figure~\ref{fig:dust_temperature} shows the luminosity of each fitted dust component ($L_\mathrm{dust}$), normalized by the rest-frame optical pseudo-blackbody luminosity ($L_{\rm pseudo-BB}$). 
\citet{lin26} report the optical pseudo-blackbody temperature of 5019~K, 5013~K, and 5558~K for J1025+1402, J1047+0739, and J1022+0841, respectively (Table~D1 of \citealt{lin26}). 
Thus, the left panel compares the relative strengths of the $T_{\rm dust}\sim90$, $300$, and $1000$~K dust templates. It does not directly measure dust mass, covering factor, or line-of-sight extinction. The $\sim2000$~K component fitted for J1022+0841 by \citet{lin26} is included in the figure. 

For reference, we estimate a characteristic dust-heating scale by assuming that the $T\sim5000$~K optical pseudo-blackbody emission illuminates the surrounding dust. Setting the luminosity of the heating source as $L_{\rm heat}=L_{\rm pseudo-BB}$, radiative equilibrium of a single dust grain gives
\begin{equation}
    r_{\rm BB} =
    \left(\frac{L_{\rm heat}}
    {16\pi\sigma_{\rm SB}T_{\rm dust}^{4}}\right)^{1/2},
    \label{eq:rheat}
\end{equation}
where $\sigma_{\rm SB}$ is the Stefan Boltzmann constant.
The upper axis of the left panel shows this quantity. It is an order-of-magnitude heating scale, not a direct size measurement or a globally energy-balanced estimate. 
Dust at $T_{\rm dust}\sim300$--$2000$~K is commonly associated with the warm and hot circumnuclear component of classical AGN tori ($r_\mathrm{BB}\sim0.01$--$1$~pc scale), whereas the $\sim90$~K component can include more extended polar, narrow-line-region, or host-galaxy dust \citep[e.g.,][]{lyu21}. J1025+1402 (red) and J1022+0841 (green) have their strongest fitted components near $300$~K, which might indicate the presence of dust around tori in the case of the classical AGNs.
On the other hand, the $\sim90$~K component is strongest in J1047+0739 (blue), which might suggest that a narrow-line region and/or host galaxy-scale dust is dominant.

The right panel of Figure~\ref{fig:dust_temperature} compares the narrow- and broad-line attenuation estimates with the H$\alpha$ FWHM. The narrow-line values are taken from \citet{lin26}. For broad lines, we derive the minimum additional SMC attenuation after adopting the maximum intrinsic ratio in our dust-free \textsc{Cloudy} grid, H$\alpha$/H$\beta\simeq13$; the inferred values are therefore model-dependent lower limits. J1025+1402 (red) and J1022+0841 (green) require $E(B-V)_{\rm broad}\gtrsim0.2$ to the dust-free \textsc{Cloudy} models. Their prominent $\sim300$--$2000$~K components (left panel of Figure~\ref{fig:dust_temperature}) are qualitatively consistent with warm circumnuclear dust contributing to the broad-line attenuation, although the MIR-emitting and attenuating dust need not be the same component. The dust-free grid reproduces J1047+0739 and hence has no positive lower limit on additional broad-line attenuation. $\sim90$~K emission of J1047+0739 (blue) may instead trace more extended dust compatible with its narrow-line attenuation ($A_{V, {\rm narrow}}\sim0.8$ for J1047+0739; \citealt{lin26}).

In addition, summing the fitted dust components relative to the pseudo-blackbody luminosity, J1022+0841 has the largest total relative warm/hot dust luminosity ($T_{\rm dust}=300$--$2000$~K) among the three local LRDs. This result is qualitatively consistent with the large broad-line attenuation inferred for J1022+0841 (green) in the right panel of Figure~\ref{fig:dust_temperature}. However, the dust luminosity traces global reprocessed emission, whereas the broad-line attenuation is line-of-sight dependent.

We note that there are some degeneracies and uncertainties in decomposing the IR luminosity of each temperature, also mentioned in \citet{lin26}. The general trend of the detection in MIR remains unchanged, but future MIR observations, such as JWST/MIRI spectroscopy for local LRDs (e.g., JWST GO-12316), will further probe these components. 

At high redshift, individual rest-frame MIR measurements remain limited (e.g., \citealt{delvecchio25, juodvbalis24, brazzini26, perez-gonzalez26}). However, stacked JWST/MIRI data show a continuum excess to $\lambda_{\rm rest}\sim3~\mu{\rm m}$ that is consistent with hot, AGN-heated dust \citep{delvecchio25}. This population-level result is consistent with dust near at least some LRD central engines, but it neither requires nor localizes the dust responsible for any individual broad-line decrement. 
In addition, the MIRI SED of the Rosetta Stone has been interpreted as containing a warm or hot dust component \citep{juodvbalis24, brazzini26}.
Taken together, the line ratio and MIR constraints support a picture in which an optically thick gas shapes the intrinsic broad hydrogen-line ratios, while additional dust attenuation may be required along some lines of sight.

\subsection{Possible Type-2-like LRDs} \label{subsec:type-2}

Although our analysis is restricted to the broad Balmer/Paschen line objects, possible narrow-line counterparts provide an external test of the sightline-dependent dust geometry. If the circumnuclear dust column increases along some viewing directions, the same underlying central engine could appear without a detectable BLR. 

In previous studies, some LRD candidates have compact morphologies and V-shaped continua but lack a detectable broad H$\alpha$ component. Using NIRSpec medium- or high-resolution spectra, \citet{zhang26} identified five such narrow-line LRDs among 32 LRDs with H$\alpha$ coverage ($\sim16\%$). These sources could be dusty, compact star-forming galaxies, but they may also include obscured AGNs. In the latter case, dust associated with circumnuclear material could attenuate BLR emission along our line of sight, making them Type-2-like counterparts of broad-line LRDs \citep[e.g.,][]{urry95}.

This interpretation is not yet unique. \citet{hviding25} find broad Balmer emission in all compact spectroscopically V-shaped point sources for which the data allowed robust broad-line measurement. The remaining cases are inconclusive due to spectral limitations. Longer-wavelength spectroscopy, including Paschen lines, will help to test whether narrow-line LRDs contain a dust-obscured BLR. If confirmed, these narrow-line LRDs may provide complementary views of central engines seen along more obscured sightlines. Together with the broad-line ratios and MIR constraints, they motivate the illustrative circumnuclear dust geometry discussed in the following subsection.

\subsection{Possible Dust Geometry} \label{subsec:dust-geometry}

%%% fig: dust temperature %%%%%%%%%%%%%
\begin{figure*}
    % \plotone{figures/fig_LRD_v1.pdf}
    \includegraphics[width=1.0\linewidth]{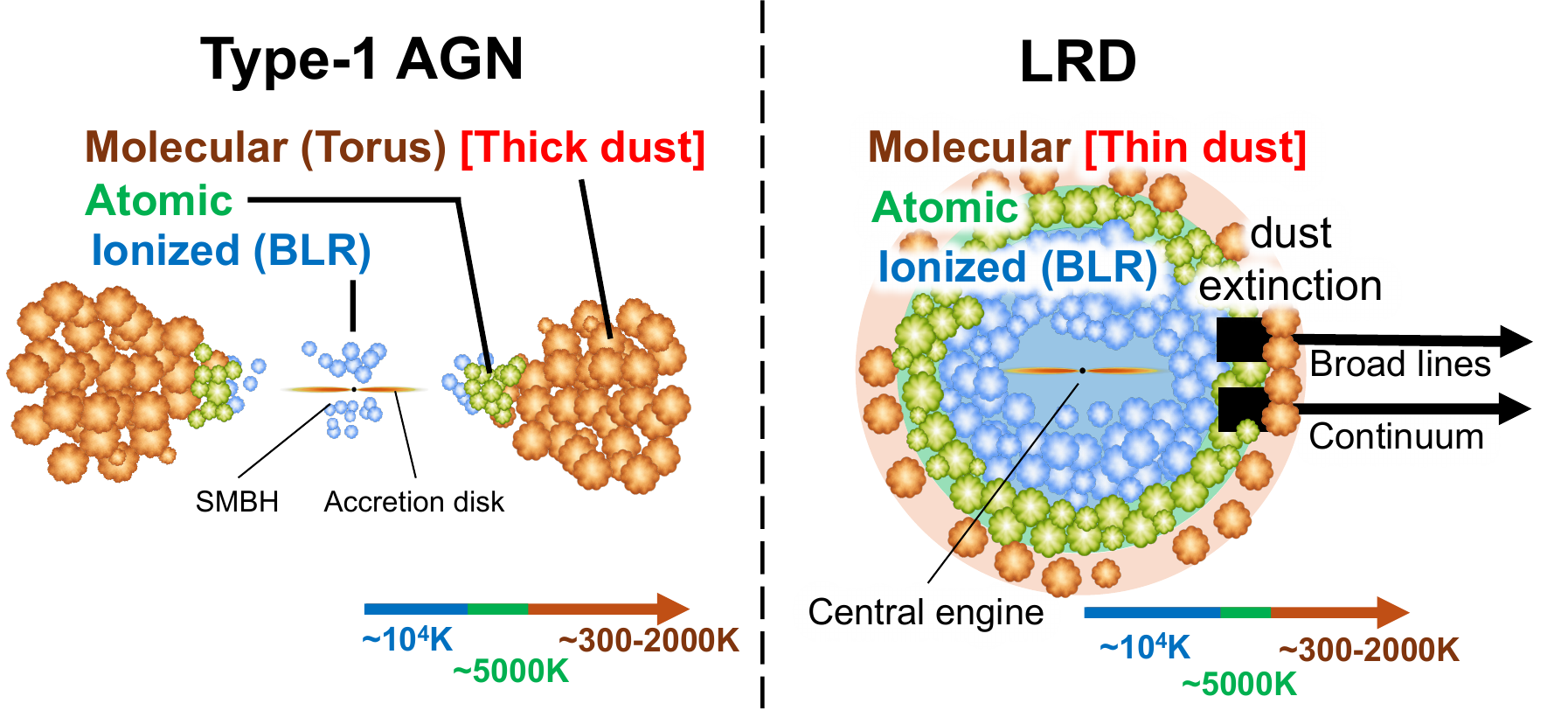}
    \caption{
    Simplified schematic view of a classical Type-1 AGN (left) and an LRD (right). The blue, green, and brown clouds denote dense ionized gas, cooler low-ionization/atomic gas, and possible molecular and dusty gas, respectively. In the LRD schematic, dust along the illustrated line of sight can attenuate the broad-line and continuum emission. The bottom arrows indicate the temperature structure ($\sim10^4$~K for ionized gas, $\sim5000$~K for atomic gas/low ionization gas, and $\sim300$--$2000$~K for molecular gas or dust temperature).
    }
\label{fig:schematic_view}
\end{figure*} 
%%%%%%%%%%%%%%%%%%%%%%%%%%%%%%%%%%%%%%

Figure~\ref{fig:schematic_view} presents an illustrative multiphase geometry of an LRD motivated by the hydrogen-line ratios and MIR constraints. Figure~\ref{fig:schematic_view} shows the simplified schematic view of a Type-1 AGN (left panel) and LRD (right panel). 
In LRDs, dense, optically thick ionized gas near the central engine (blue in Figure~\ref{fig:schematic_view}) can produce non-Case-B broad hydrogen-line ratios, while warm or hot circumnuclear dust (brown in Figure~\ref{fig:schematic_view}) may additionally attenuate the broad-line-emitting gas along some lines of sight. 
The optical continuum temperature inferred for LRDs ($\sim5000~{\rm K}$; e.g., \citealt{degraaff26-LRD}) exceeds the dust sublimation temperature ($\sim1500$--$2000$~K; e.g., \citealt{barvainis87, baskin18}), implying that dust is unlikely to survive in the innermost line-emitting region. 
As the incident radiation field decreases and shielding increases with distance from the central engine, conditions may favor cooler, low-ionization, molecular, and dusty gas ($T_{\rm dust}\sim300$--$2000~{\rm K}$) phases. The narrow-line emission may probe more extended regions and different sightlines and can therefore experience less attenuation. 

Additional spectral features support the presence of such multiphase/stratified gas structures in at least some LRDs \citep[e.g.,][]{lin26, wang26, kokorev26, torralba26}. For example, the local LRD J1025+1402 shows strong Ca\,{\sc ii} triplet, Na~D, and K\,{\sc i} absorption, suggesting cool, metal-enriched, low-ionization/atomic gas \citep[green in Figure~\ref{fig:schematic_view};][see also \citealt{greene26} for the Ca\,{\sc ii} triplet of high-redshift LRDs]{lin26}. 
\citet{wang26} reported a rest-frame $\sim1.4~\mu{\rm m}$ absorption characteristic consistent with H$_2$O absorption in two LRDs, indicating a cool ($T\lesssim3000$~K), molecular gas component \citep[brown in Figure~\ref{fig:schematic_view}; see also][]{liu26}. The LRD schematic view qualitatively resembles the stratified Type-1 AGNs (left panel of Figure~\ref{fig:schematic_view}), mainly determined by the distance from the central engine and hence gas/dust temperature. If its circumnuclear gas redistributes angular momentum, an LRD could develop a flatter, classical AGN-like dusty structure. 

The circumnuclear dust in LRDs may represent a lower-column-density counterpart of the classical AGN torus, whose dusty clouds are optically thick at rest-frame $V$ band. This proposed contrast is illustrated by the labels ``thick dust'' and ``thin dust'' in Figure~\ref{fig:schematic_view}. The latter does not necessarily imply that the dust is optically thin at a particular wavelength. 

We refer to this putative circumnuclear dust component as a ``proto-torus,'' while emphasizing that its geometry, column density, and evolutionary relation to classical AGN tori remain unconstrained. Among the four high-redshift LRDs with luminous broad H$\alpha$ ($\log(L_{\rm H\alpha,broad}/{\rm erg~s^{-1}})>43.5$), one has H$\alpha$/H$\beta>13$ (Figure~\ref{fig:broadHaHb_Haluminosity}). Since our fiducial \textsc{Cloudy} total-emission models cannot reproduce such a large decrement, this object could require additional dust attenuation. The observed incidence ($1/4$) is suggestive, but cannot yet be converted directly into a dust-covering factor due to the small, heterogeneous sample. Larger homogeneous samples with Balmer and Paschen spectroscopy will be needed to constrain the geometry of circumnuclear dust in LRDs. 
Cold-dust ($\sim30$~K) constraints for high-redshift LRDs from the Atacama Large Millimeter/submillimeter Array (ALMA) \citep[e.g.,][]{casey25, setton25, xiao25} probe a different component and do not by themselves rule out this hot/warm dust scenario (i.e., Dust-poor on the galaxy scale, but dust is in LRDs). 
Complementary constraints on a local LRD are the aim of the approved ALMA Cycle~13 program (2026.1.00880.S; PI: T. Kiyota), which targets the Band~7 continuum in ``The Egg''.

An additional speculative possibility is in-situ dust formation within the dense gas surrounding the central engine. \citet{naidu26} proposed that LRDs may share similarities with Type IIn supernovae, in which shocks within a dense circumstellar envelope can produce favorable conditions for dust formation \citep[e.g.,][]{Shahbandeh25}. If analogous processes operate in LRDs, they could provide a pathway for producing circumnuclear dust around the broad-line-emitting gas \citep[see also][]{ashall26}. Although this scenario is not tested with the present data, the dust-production mechanism itself is an intriguing avenue for future studies of LRD central engines.

\section{Summary and Conclusion} \label{sec:summary} 

We investigate broad Balmer and Paschen line ratios in a sample of 20 LRDs, comprising 15 low-redshift ($z=0.1$--$0.9$) and five high-redshift ($z=2.3$--$7.0$) sources. Our sample requires both broad H$\alpha$ and broad H$\beta$ to be detected at
${\rm S/N}\geq5$. By comparing the measured line ratios with dust-free \textsc{Cloudy} calculations, we assess the relative roles of optically thick gas and dust attenuation in shaping the broad hydrogen-line emission. Our main results are summarized as follows. 

\begin{enumerate}
\item Broad H$\alpha$/H$\beta$ ratios ($\sim6$--$30$) are generally larger than the corresponding narrow-line ratios ($\sim3$--$5$) and the Case B value ($=2.87$ for $n_e=10^2~{\rm cm^{-3}}$ and $T_e=10^4$~K). In contrast, the narrow H$\alpha$/H$\beta$ ratios are broadly consistent with the Case B recombination value. This broad-narrow contrast indicates that the physical conditions and/or attenuation affecting the broad-line-emitting gas differ from those affecting the narrow-line-emitting gas.

\item The two local LRDs with reliable broad Pa$\alpha$, Pa$\beta$ measurements (J1022+0841 and J1047+0739) have Pa$\alpha$/Pa$\beta$ ratios ($\sim1.5$) below the adopted Case B value ($=2.0$). Because a foreground attenuation screen would increase this ratio, the Paschen measurements support non-Case-B hydrogen-line formation processes in an optically thick, high-density gas.

\item The dust-free \textsc{Cloudy} grid (assuming plane-parallel geometry and adopting total line intensity outputs) reproduces elevated broad Balmer decrements through radiative-transfer and collisional effects, with the largest H$\alpha$/H$\beta$ ratios reaching $\sim13$ at $\log(n_{\rm H}/{\rm cm^{-3}})\sim10$ and $\log(\Phi/{\rm cm^{-2} ~s^{-1}})\sim18$. 
Broad H$\alpha$/H$\beta$ ratios above this model would require additional differential attenuation, or physical conditions not included in the present grid. 

\item
Among the three LRDs with joint Balmer--Paschen constraints for objects with Paschen line measurements, J1047+0739 can be approximately reproduced by dust-free dense-gas models. J1022+0841 and Rosetta Stone remain outside the dust-free grid and require additional attenuation within the model framework.

\item 
For sources without usable Paschen constraints, we apply the maximum H$\alpha$/H$\beta\sim13$ as a diagnostic of the additional attenuation.
Two high-redshift and eight low-redshift LRDs in the sample have broad H$\alpha$/H$\beta$ ratios above the maximum value of the \textsc{Cloudy} grids ($\mathrm{H\alpha/H\beta}>13$).
If these extreme decrements are interpreted as additional dust attenuation, they imply that dust preferentially attenuates the broad-line emission along some lines of sight with $E(B-V)\gtrsim0.2$.

\item The warm/hot dust components inferred from WISE observations ($\lambda_{\rm rest}\sim3$--$20~\mu$m) of local LRDs and the stacked rest-frame MIR excess ($\lambda_{\rm rest}=3$~$\mu$m) reported for high-redshift LRDs are consistent with hot/warm dust ($\sim300$--$1000$~K) near central engines of LRDs. We note that MIR-emitting dust need not be identical to the material responsible for broad-line attenuation. Together, the line-ratio and MIR constraints are consistent with dense, optically thick line-forming gas surrounded by, or viewed through, a low-column-density counterpart of the classical thick AGN torus, which we refer to as a ``proto-torus.''
\end{enumerate}

Future JWST/MIRI spectroscopy (e.g., GO-10116, 10445, 11376, 12312, 12396, and future observations) will expand the sample of LRDs with Paschen series coverage, especially at higher redshifts. These data will provide key diagnostics of the physical conditions in the central engines of LRDs, including the non-Case B conditions and the existence of dust. Deeper MIR observations, such as the PRobe for-Infrared Mission for Astrophysics (PRIMA; \citealt{jason25}), will also provide constraints on the dust continuum emission of LRDs, especially from warm/hot dust.

\begin{acknowledgments} \label{ack}

The authors thank the anonymous referee for the careful reading and valuable comments that improved the manuscript.
The authors thank Madisyn Brooks, Martin Gaskell, Anna de Graaff, Ignas Juod{\v{z}}balis, Vasily Kokorev, Mitsuru Kokubo, Rohan P. Naidu, Giorgos P. Nikopoulos, and Taketo Yoshida for valuable comments and discussions. We thank Anna de Graaff et al. for the public release of the JWST LRD catalog. We thank Xiaojing Lin et al. for conducting the local LRD studies. 

This work is based in part on observations made with the NASA/ESA/CSA James Webb Space Telescope. 
The JWST data were obtained from the Mikulski Archive for Space Telescopes at the Space Telescope Science Institute, which is operated by the Association of Universities for Research in Astronomy, Inc., under NASA contract NAS 5-03127 for JWST. 
The specific observations analyzed can be accessed via \dataset[doi: 10.17909/fehq-p836]{https://doi.org/10.17909/fehq-p836}.
These observations are associated with JADES GTO-1180, JADES GTO-1181 (PI: D.~J. Eisenstein), JADES GTO-1286 (PI: N. Lützgendorf), CEERS ERS-1345 (PI: Steven L. Finkelstein), DeepDive GO-3567 (PI: Francesco Valentino), GO-4106 (PI: Erica Nelson), and RUBIES GO-4233 (PI: A. de Graaff and G. Brammer). The authors acknowledge these teams for developing their observing programs.

The JWST/NIRSpec data products presented herein were retrieved from the Dawn JWST Archive (DJA). DJA is an initiative of the Cosmic Dawn Center (DAWN), which is funded by the Danish National Research Foundation under grant DNRF140. We thank DAWN for providing the reduced JWST/NIRSpec data.

T.K. acknowledges support by KAKENHI (26KJ1232) through Japan Society for the Promotion of Science (JSPS). T.K. was partially supported by the SOKENDAI Student Dispatch Program (2026), The Graduate University for Advanced Studies, SOKENDAI. T.K. was also partially supported by the Overseas Travel Fund for Students (2026) of the Astronomical Science Program, The Graduate University for Advanced Studies, SOKENDAI. 
M.O. acknowledges the support from the World Premier International Research Center Initiative (WPI Initiative), MEXT, Japan, the joint research program of the Institute for Cosmic Ray Research (ICRR), the University of Tokyo, and KAKENHI (21H04467, 25H00674) through Japan Society for the Promotion of Science (JSPS). 
H.Y. acknowledges support by KAKENHI (25KJ0832) through Japan Society for the Promotion of Science (JSPS). 
Y.K. acknowledges support by KAKENHI (26KJ0960) through JSPS, JSR Fellowship, and FoPM, WINGS Program, the University of Tokyo. 
M.N. is supported by JSPS KAKENHI Grant No. 25KJ0828. 
Y.H. acknowledges support from the Japan Society for the Promotion of Science (JSPS) Grant-in-Aid for Scientific Research (24H00245) and the JSPS International Leading Research (22K21349). 
Y.N. acknowledges Flatiron Research Fellowship. The Flatiron Institute is a division of the Simons Foundation.

Data analysis was carried out in part on the Multiwavelength Data Analysis System operated by the Astronomy Data Center (ADC), National Astronomical Observatory of Japan. The English writing in this paper has been improved with the help of OpenAI ChatGPT and Grammarly, but the software does not generate sentences from scratch. 

\end{acknowledgments}

\facilities{JWST/NIRSpec}

\software{Astropy \citep{astropy:2013, astropy:2018, astropy:2022},  
          \textsc{Cloudy} \citep{Gunasekera23}, 
          emcee \citep{mackey13}, 
          LMFIT \citep{newville25}, 
          Matplotlib \citep{hunter07},
          NumPy \citep{harris20}, 
          pandas \citep{mckinney2010data}
          PyNeb \citep{luridiana15},
          SciPy \citep{virtanen20}, 
          }

\appendix
\twocolumngrid
\restartappendixnumbering

\section{WISE Photometry of the Three Local LRDs}
\label{app:wise-photometry}

We compile the observed-frame WISE W1-W4 flux densities of the three local LRDs in Table~\ref{tab:local-lrd-wise}. These fluxes are taken from the Legacy Surveys DR10 catalog \citep{dey19} based on the unWISE images \citep{schlafly19}, which are also used by \citet{lin26}. These measurements are provided as supplementary information and are not used directly in our analysis. The dust-component luminosities discussed in Section~\ref{sec:mid-infrared-constraints} are adopted from \citet{lin26}. We do not perform a new fit to the WISE photometry.

\begin{deluxetable}{lcccc}
\tabletypesize{\small}
\tablewidth{0pt}
\tablecaption{WISE flux densities of the three SDSS local LRDs.
\label{tab:local-lrd-wise}}
\tablehead{
\colhead{Object} & \colhead{W1} & \colhead{W2} &
\colhead{W3} & \colhead{W4} \\
\colhead{} & \colhead{(mJy)} & \colhead{(mJy)} &
\colhead{(mJy)} & \colhead{(mJy)}
}
\startdata
J1047+0739 & $0.2875\pm0.0025$ & $0.7066\pm0.0059$ &
$6.76\pm0.15$ & $41.3\pm2.5$ \\
J1025+1402 & $0.2914\pm0.0025$ & $0.3874\pm0.0053$ &
$2.58\pm0.15$ & $5.5\pm2.5$ \\
J1022+0841 & $0.1618\pm0.0023$ & $0.2630\pm0.0051$ &
$1.59\pm0.14$ & $8.4\pm2.5$ \\
\enddata
\tablecomments{Observed-frame flux densities $f_{\nu}$ from the Legacy Surveys DR10 catalog \citep{dey19}, which provides unWISE photometry \citep{schlafly19}. Uncertainties are formal catalog $1\sigma$ uncertainties. No Galactic-extinction correction or $K$-correction has been applied.}
\end{deluxetable}

\restartappendixnumbering
\section{Cloudy Model Grid Results} 
\label{appendix:cloudy-model} 

\begin{figure*}
    \plotone{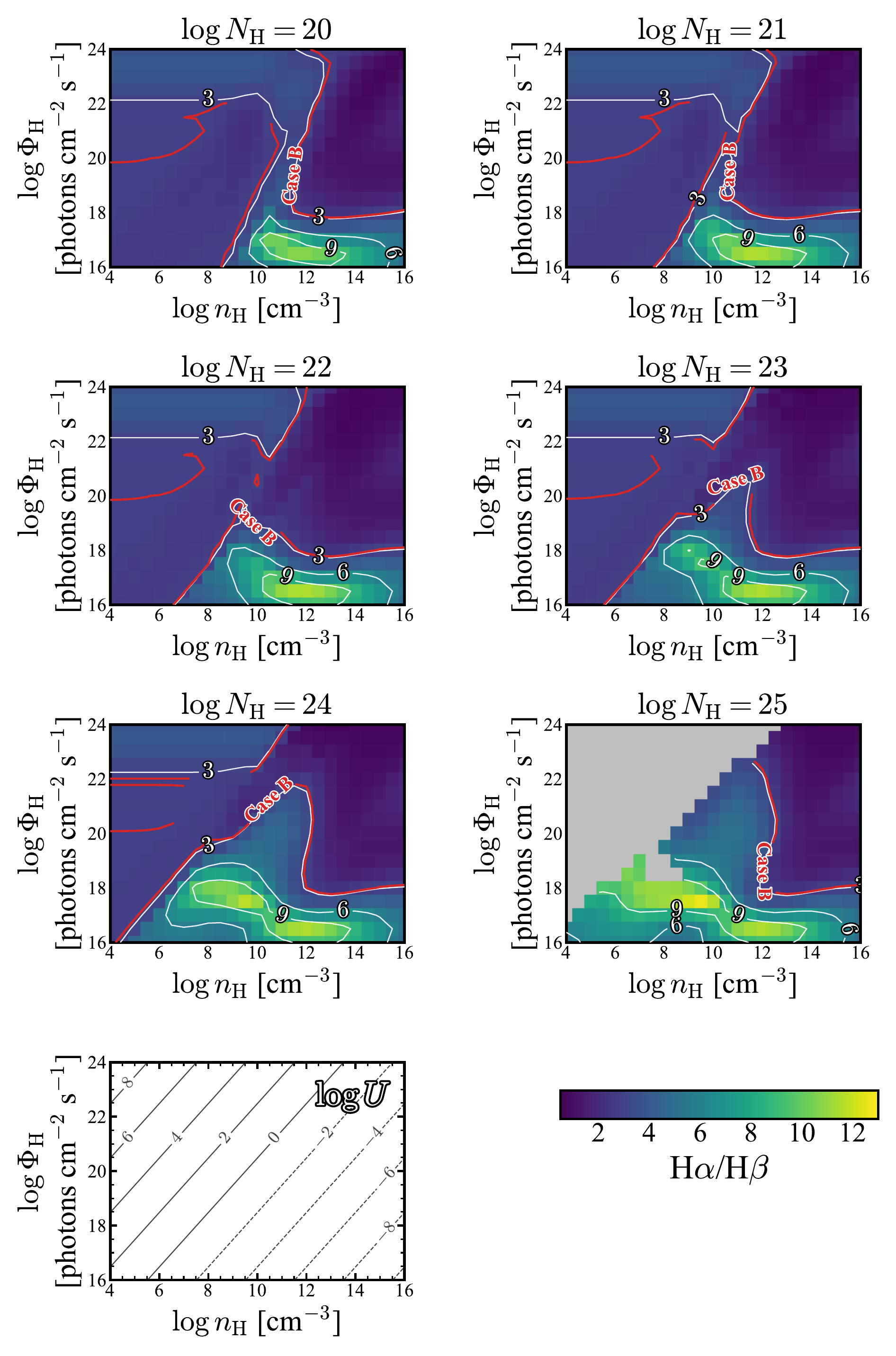}
    \caption{ 
    The model grids of the $\Phi({\rm H})$--$n_{\rm H}$ plane for the H$\alpha$/H$\beta$ (see also \citealt{korista04, Schnorr-Muller16}). The color shows the H$\alpha$/H$\beta$ ratio obtained from the \textsc{Cloudy} modeling. The white contours show H$\alpha$/H$\beta$ ratios of 3, 6, and 9. The red contours show the Case B intrinsic value ($=2.87$). Each panel represents the different column density ($\log(N_{\rm H}/{\rm cm^{-2}})=20$--$25$). The gray region is where the \textsc{Cloudy} calculation fails to complete due to nonphysical conditions. The bottom left panel shows the corresponding ionization parameter $\log U$ for each $\log\Phi({\rm H})$ and $\log n_{\rm H}$. 
    }
\label{fig:cloudy_grids_HaHb}
\end{figure*}

\begin{figure*}
    \plotone{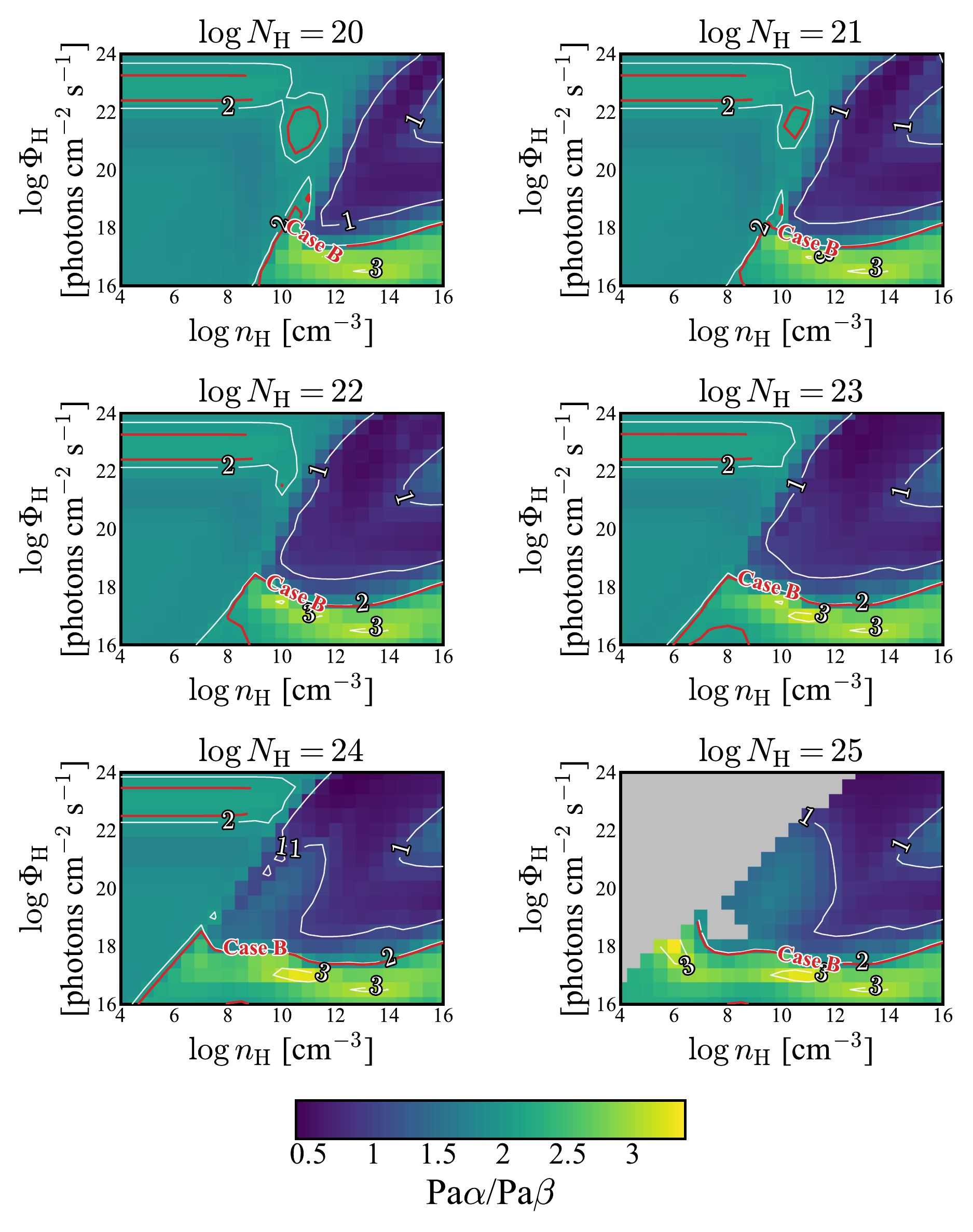}
    \caption{
    Same as in Figure~\ref{fig:cloudy_grids_HaHb}, but for Pa$\alpha$/Pa$\beta$. 
    }
\label{fig:cloudy_grids_PaaPab}
\end{figure*}

\begin{figure*}
    \plotone{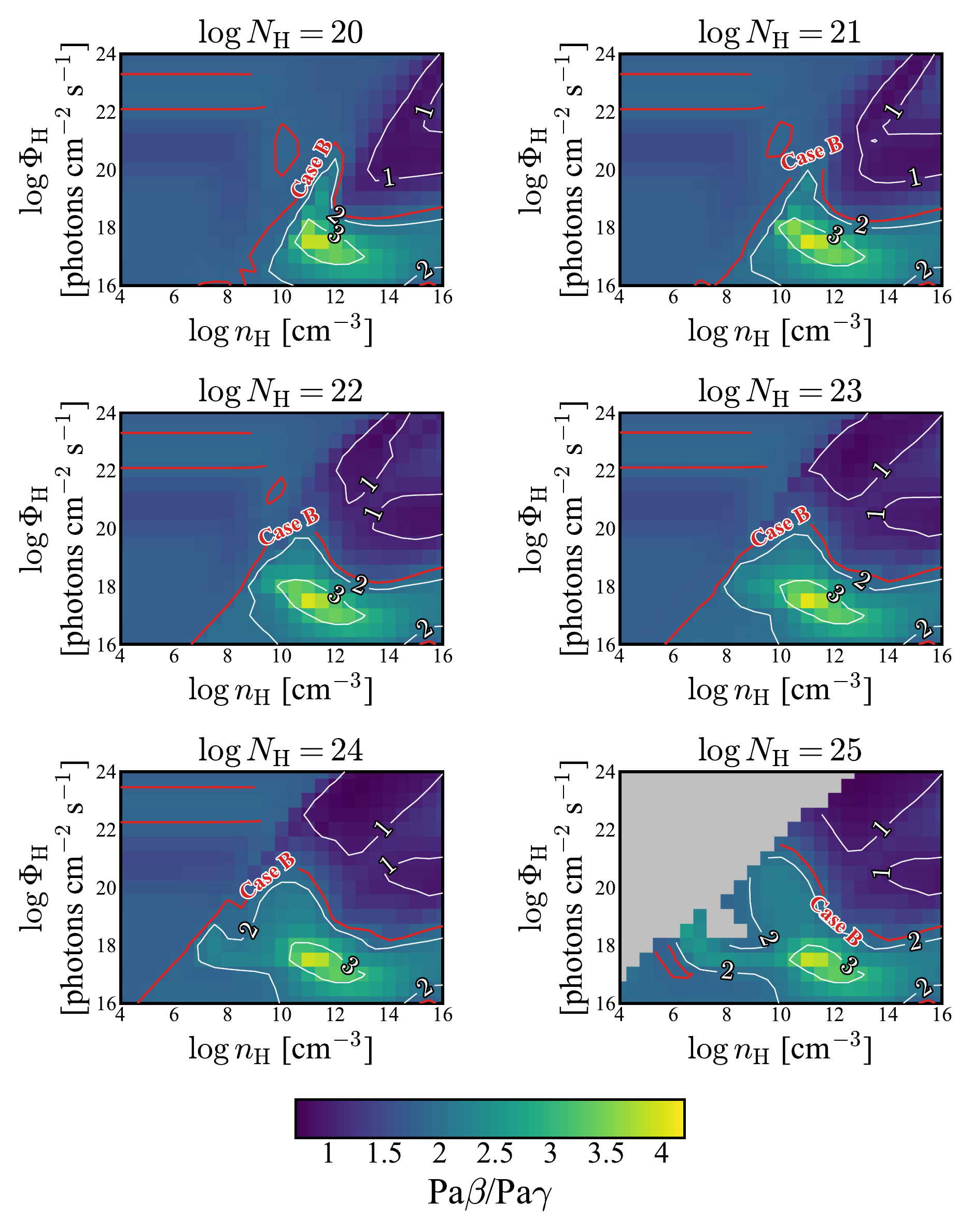}
    \caption{
    Same as in Figure~\ref{fig:cloudy_grids_HaHb}, but for Pa$\beta$/Pa$\gamma$. 
    }
\label{fig:cloudy_grids_PabPag}
\end{figure*}

We show the \textsc{Cloudy} modeling results obtained from ``total'' emission line intensity outputs in the $\Phi({\rm H})$-$n_{\rm H}$ plane for H$\alpha$/H$\beta$, Pa$\alpha$/Pa$\beta$, and Pa$\beta$/Pa$\gamma$ in Figure~\ref{fig:cloudy_grids_HaHb}, Figure~\ref{fig:cloudy_grids_PaaPab}, and Figure~\ref{fig:cloudy_grids_PabPag}, respectively.

\restartappendixnumbering
\section{Sensitivity to Geometry and Directional Line Output}
\label{appendix:cloudy-model-output} 

\begin{figure*}
    \plotone{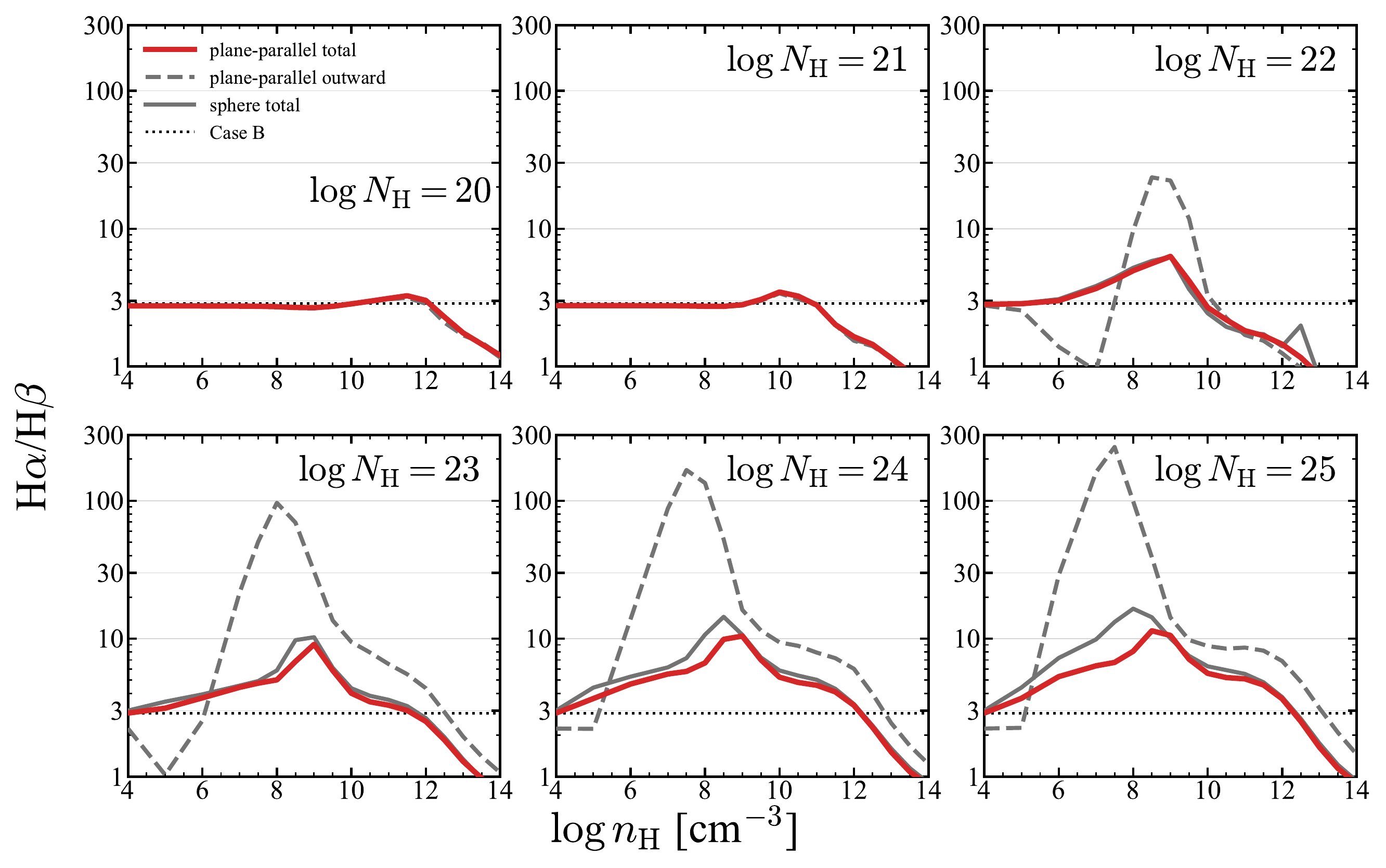}
    \caption{
    Cloudy predictions for the H$\alpha$/H$\beta$ line ratio as a function of hydrogen density, $\log n_{\rm H}$, at $\log U=-1.48$ (the grid value closest to $-1.5$), metallicity of $Z/Z_\odot=0.1$. The panels show the hydrogen column density, $\log N_{\rm H}=20$--$25$ from upper left to lower right. The red solid curves show the plane-parallel total line ratio, the gray dashed curves show the plane-parallel outward component, computed as total minus inward emission, and the gray solid curves show the spherical-model total line ratio. The black dotted line marks the Case B recombination value, H$\alpha$/H$\beta=2.87$. 
    }
\label{fig:cloudy_comparison_HaHb}
\end{figure*}

\begin{figure*}
    \plotone{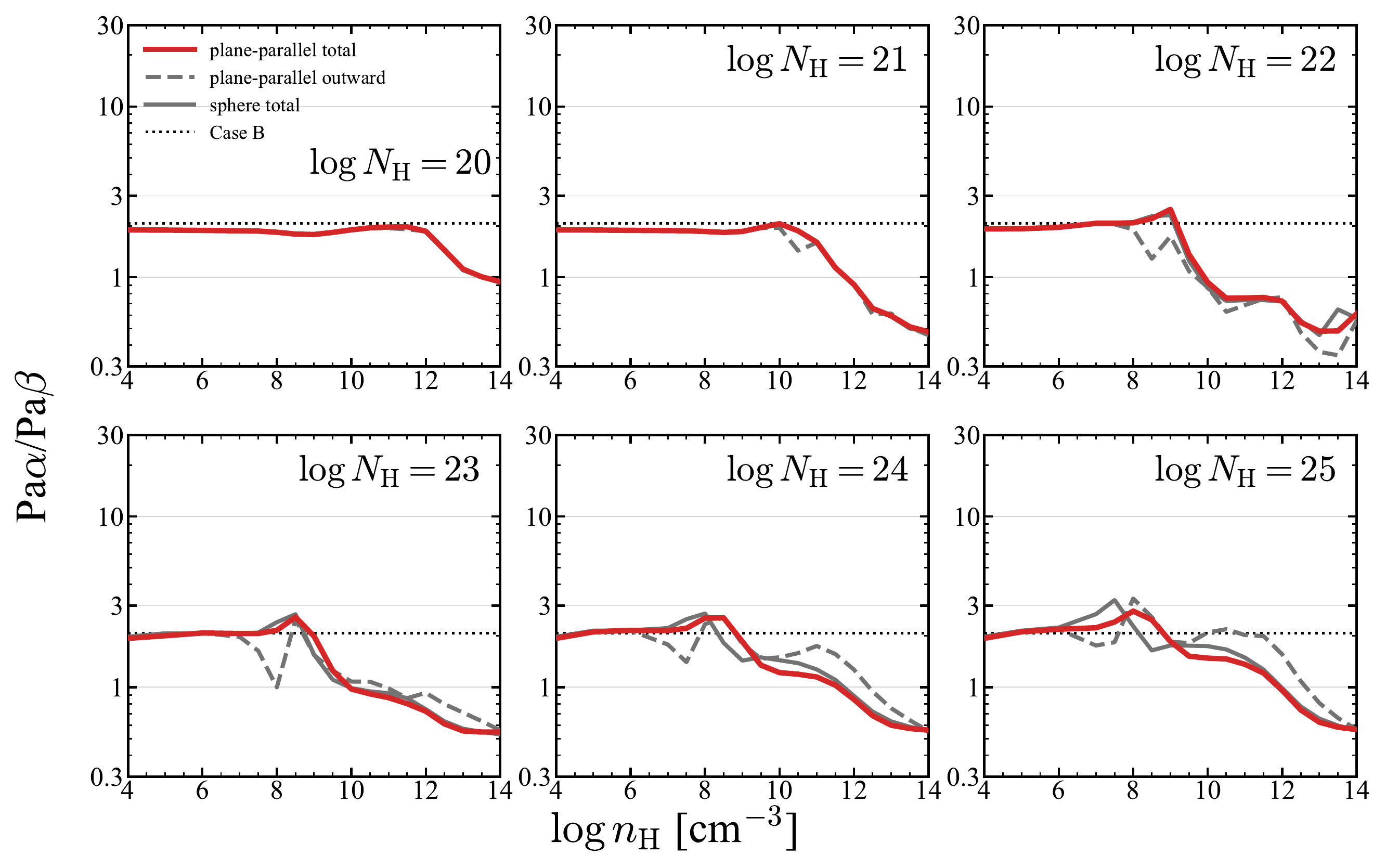}
    \caption{
    Same as in Figure~\ref{fig:cloudy_comparison_HaHb}, but for Pa$\alpha$/Pa$\beta$.
    }
\label{fig:cloudy_comparison_PaaPab}
\end{figure*}

\begin{figure*}
    \plotone{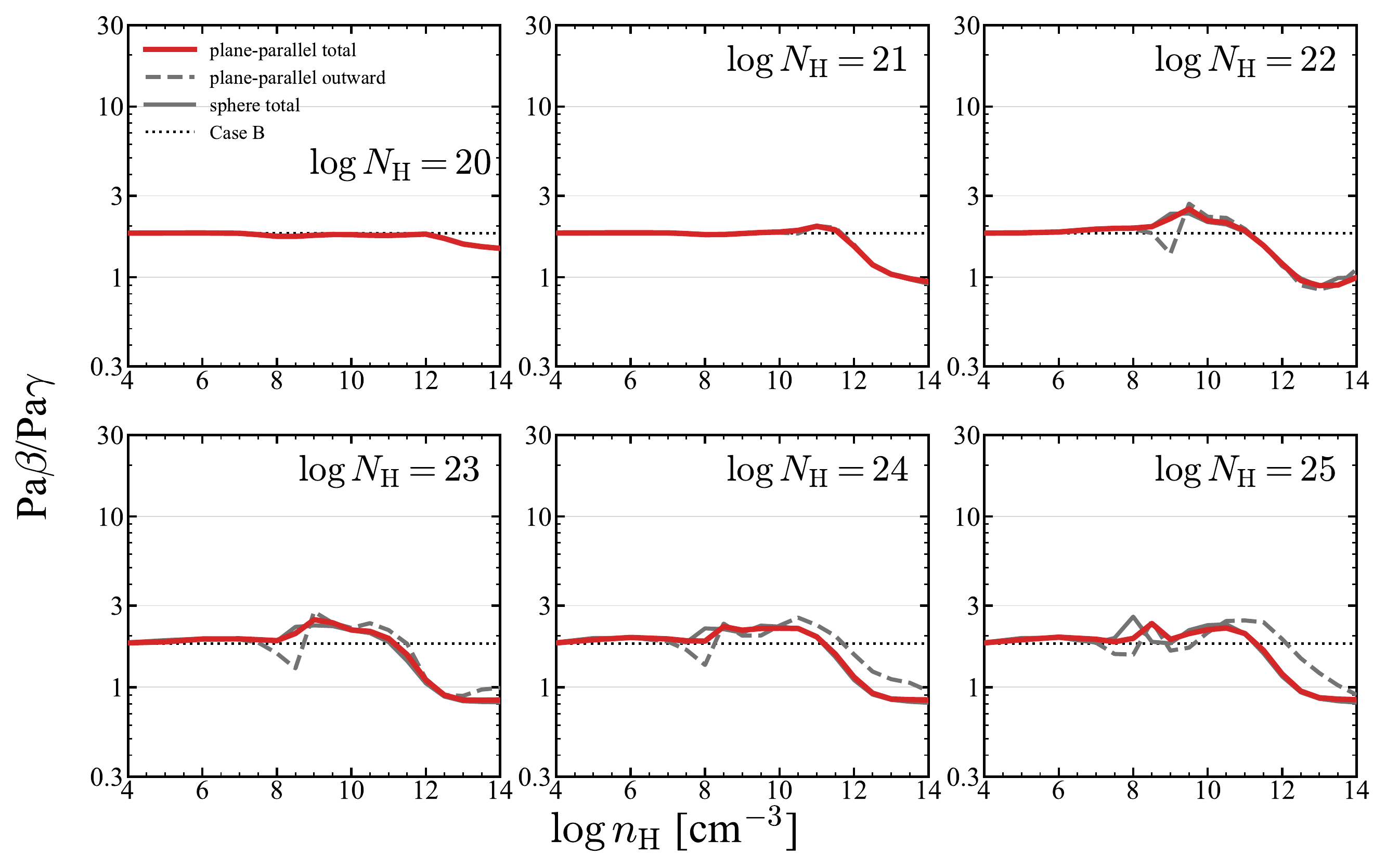}
    \caption{
    Same as in Figure~\ref{fig:cloudy_comparison_HaHb}, but for Pa$\beta$/Pa$\gamma$.
    }
\label{fig:cloudy_comparison_PabPag}
\end{figure*}

The fiducial calculations adopt the total line intensity from open plane-parallel slabs. To test this choice, we compare the total ($I_\mathrm{tot}$; usual \textsc{Cloudy} output) and the outward-only ($I_\mathrm{out}$) plane-parallel outputs with the total emission from the closed spherical models (\textsc{Cloudy} command \texttt{sphere}). For the plane-parallel models, we calculate the shielded-face emission as $I_{\rm out}=I_{\rm tot}-I_{\rm in}$, where $I_{\rm in}$ leaves the illuminated face toward the ionizing source. Figures~\ref{fig:cloudy_comparison_HaHb}, \ref{fig:cloudy_comparison_PaaPab}, and \ref{fig:cloudy_comparison_PabPag} show the resulting hydrogen-line ratios of H$\alpha$/H$\beta$, Pa$\alpha$/Pa$\beta$, and Pa$\beta$/Pa$\gamma$, respectively. We show the values at $\log U=-1.48$ for reference. 
The outward-only models provide the most extreme values, reaching H$\alpha$/H$\beta>100$ near $\log(n_{\rm H}/{\rm cm^{-3}})\sim8$ \citep[see also][]{yan26}. Thus, the Balmer decrement alone might not exclude a dust-free configuration dominated by shielded-face ``outward'' emission. 

We nevertheless adopt the total plane-parallel emission as the fiducial reference. The LRD broad lines are expected to arise from gas distributed around the central engine rather than from clouds viewed exclusively from its shielded face. In several optically thick models, the inward component accounts for a substantial fraction of the intrinsic line emission. Selecting only the outward component might therefore discard a large fraction of the line-emitting gas in an angle-integrated description. 
Moreover, total-emission models with a large covering factor reproduce the observed $L_{\rm H\alpha,\,broad}$--$L_{\rm bol}$ scaling relation \citep{yanagisawa26}. Retaining only one directional component would reduce the predicted H$\alpha$ luminosity. 

\begin{figure*}
    \includegraphics[width=1.0\linewidth]{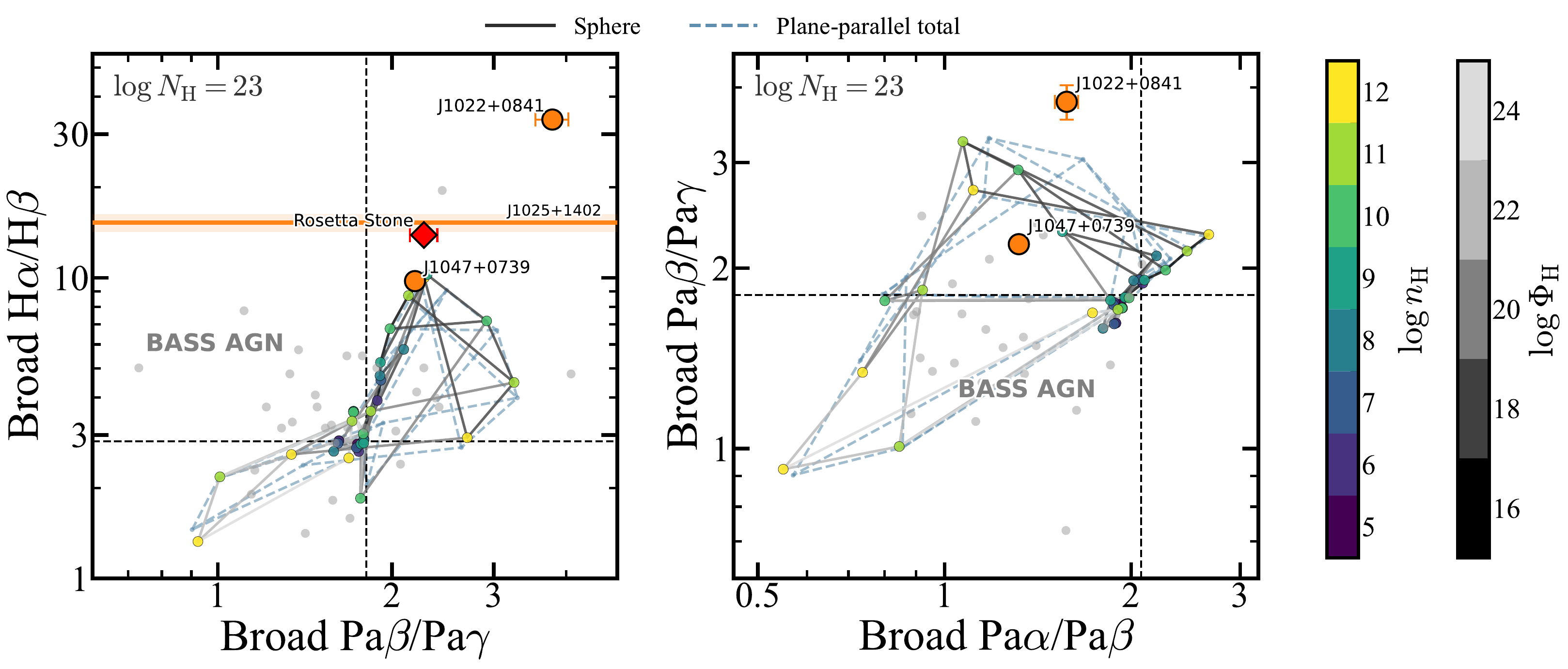}
    \caption{
    Same as in Figure~\ref{fig:HaHb_PabPag_cloudy} (left) and Figure~\ref{fig:PabPag_PaaPab_cloudy} (right), but for the model grids of the sphere geometry. For reference, the model grids of the plane-parallel geometry are shown as the blue dashed lines. The other symbols are the same as in Figure~\ref{fig:HaHb_PabPag_cloudy} and Figure~\ref{fig:PabPag_PaaPab_cloudy}. 
    }
\label{fig:cloudy_sphere}
\end{figure*}

In the closed spherical calculations, most of the total hydrogen-line intensity is assigned to the inward component in \textsc{Cloudy}. In this geometry, however, inward emission is included in the closed-shell
calculation rather than being discarded as in the outward-only slab limit.
The total spherical and total plane-parallel models also show broadly similar regions of the H$\alpha$/H$\beta$--Pa$\beta$/Pa$\gamma$ and Pa$\alpha$/Pa$\beta$--Pa$\beta$/Pa$\gamma$ planes as shown by blue dashed lines in Figure~\ref{fig:cloudy_sphere}. 

Finally, the outward-only models producing the largest Balmer decrements do not simultaneously reproduce the observed Balmer and Paschen ratios at the same model parameters. In particular, the Pa$\beta$/Pa$\gamma$ ratio of J1022+0841 remains outside the dust-free model range and requires additional attenuation ($E(B-V)\sim1$ in Figure~\ref{fig:dust_temperature}). We therefore treat the outward-only calculations as a directional upper envelope and the total plane-parallel calculations as the fiducial global-emission reference. The inference of additional dust attenuation from the joint Balmer and Paschen ratios would not be consequently driven solely by the adopted geometry or directional output.

\restartappendixnumbering
\section{SDSS High Balmer Decrement AGN Candidates} 
\label{appendix:sdss_high_HaHb}

\begin{figure*}
    \includegraphics[width=1.0\linewidth]{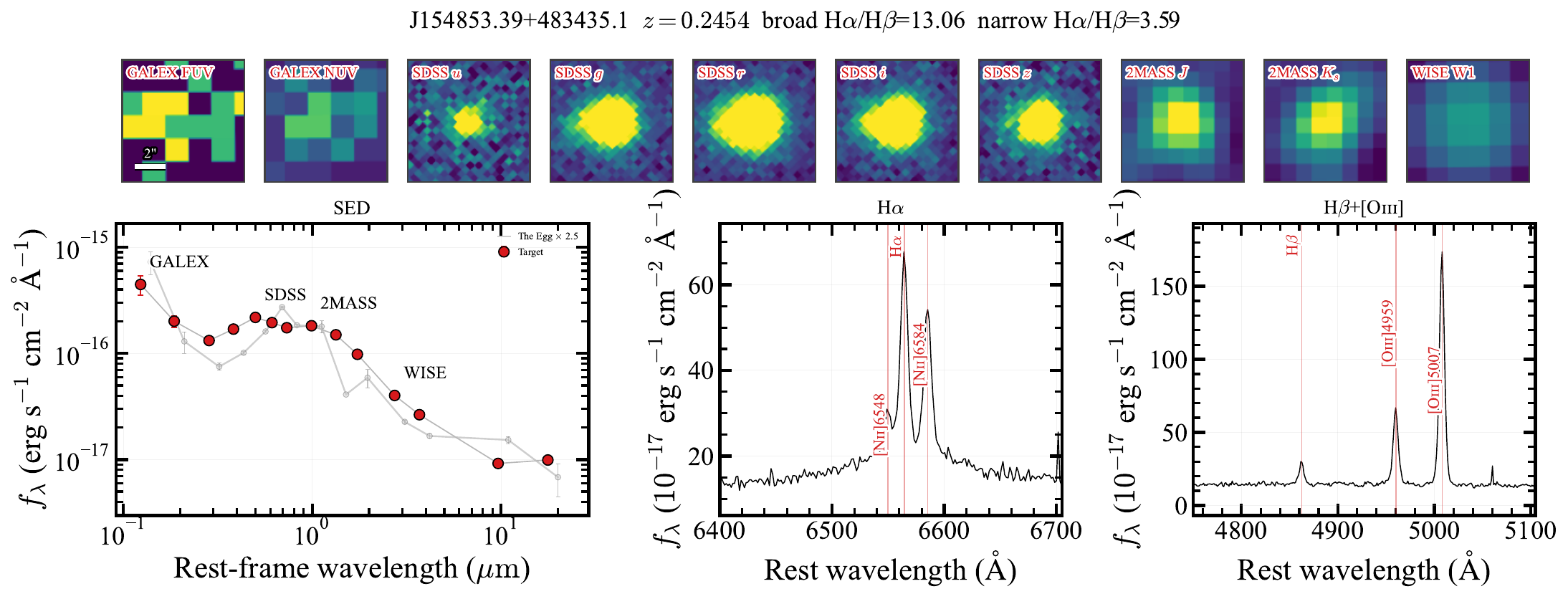}
    \includegraphics[width=1.0\linewidth]{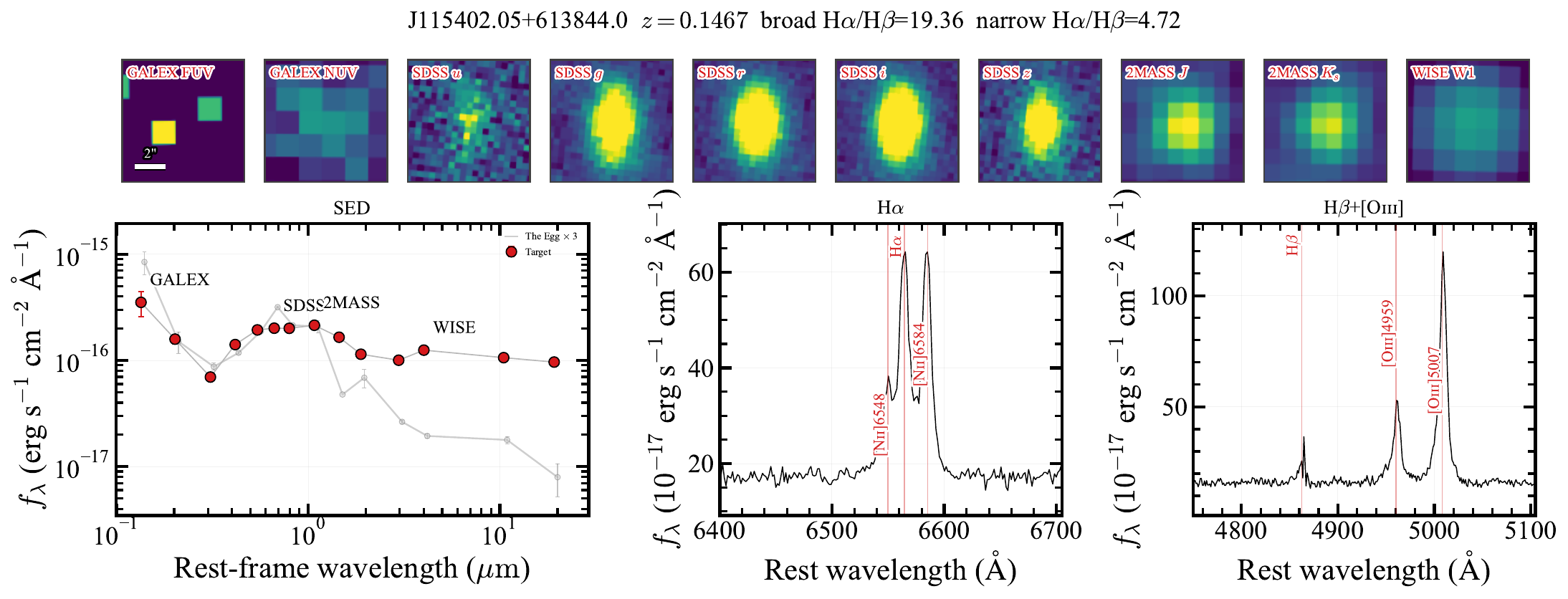}
    \includegraphics[width=1.0\linewidth]{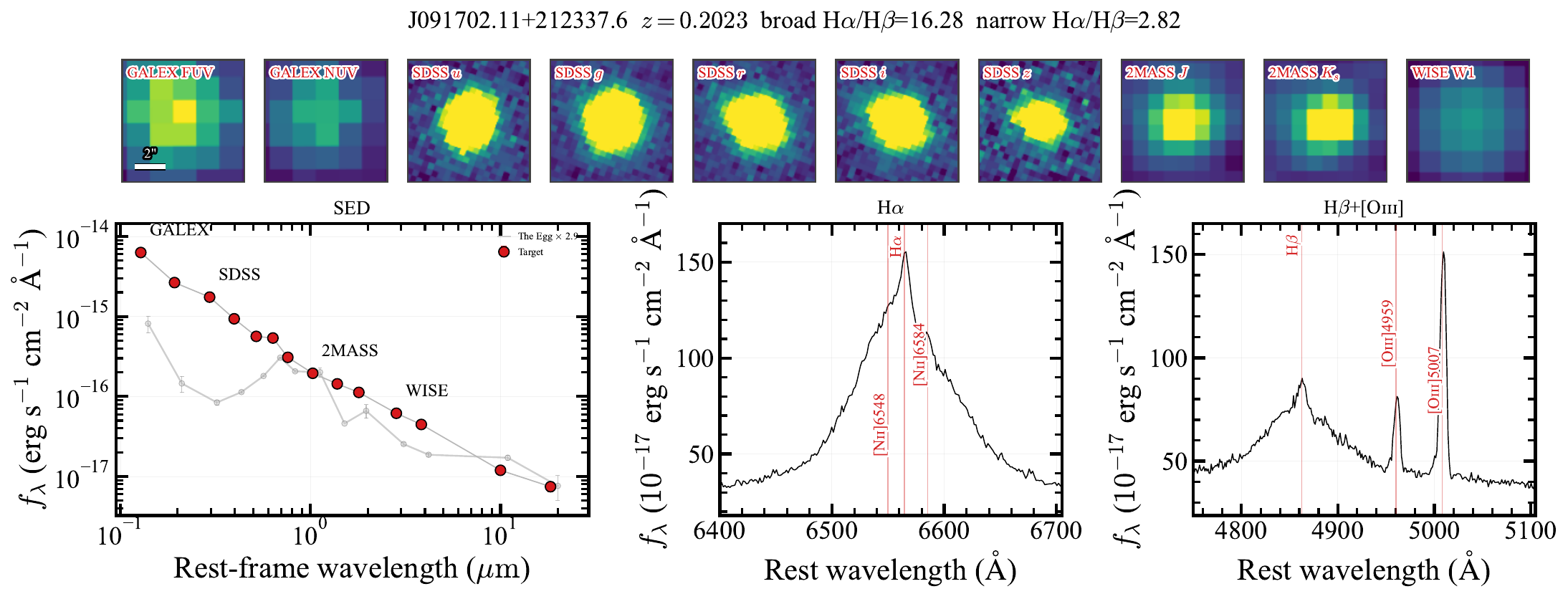}
    \caption{
    The representative SDSS extreme-decrement AGN candidates. For each object, the top row shows the imaging cutouts from the GALEX, SDSS, 2MASS, and WISE. The lower-left panel shows the SED. The red circles show the target photometry, while the gray symbols and curves show the SED of the local LRD ``The Egg'' (J1025+1402) scaled by the factor indicated in each panel. The lower middle and right panels show the SDSS spectra around the H$\alpha$+\nii\ and H$\beta$+\oiii\ complexes, respectively. The vertical red lines mark the expected wavelengths of the indicated emission lines. The source name, redshift, and the broad and narrow H$\alpha$/H$\beta$ ratios from \citet{liu19} are listed above each panel.
    }
\label{fig:sdss_highHaHb}
% J154853.39+483435.1 (V-shaped SED example)
% J091702.11+212337.6 (blue SED) 
% J115402.05+613844.0 (V-shaped SED but strong WISE)
\end{figure*}

We find 167 SDSS AGNs at $z<0.35$ that satisfy broad H$\alpha$/H$\beta>13$, ${\rm S/N}({\rm broad~H}\alpha)>3$, and ${\rm S/N}({\rm broad~H}\beta)>3$ in the catalog of \citet{liu19} (see also the gray points of Figures~\ref{fig:broadHaHb_Haluminosity} and \ref{fig:broadHaHb_narrowHaHb}). We refer to these objects as extreme-decrement AGN candidates. To provide a qualitative comparison with LRDs, we compile the Galaxy Evolution Explorer (GALEX; \citealt{martin05}), SDSS, Two Micron All Sky Survey (2MASS; \citealt{skrutskie06}), and WISE photometry listed by \citet{liu19}, inspect WISE multi-epoch photometry, and visually examine the SDSS spectra. Figure~\ref{fig:sdss_highHaHb} shows representative examples of the SED and SDSS spectra. The candidate sample shows various features, including V-shaped SEDs with weak WISE emission like LRDs (top panel of Figure~\ref{fig:sdss_highHaHb}), V-shaped SEDs with stronger MIR emission than that of ``The Egg'' (middle panel of Figure~\ref{fig:sdss_highHaHb}), and sources without a V-shaped SED (bottom panel of Figure~\ref{fig:sdss_highHaHb}). 

Some parts of objects show compact and V-shaped SEDs qualitatively similar to those of LRDs. Weak WISE emission relative to the optical continuum looks consistent with a weak warm/hot-dust component compared with dusty galaxies such as dust-obscured galaxies (DOGs), or ALMA dust continuum sources \citep[e.g.,][]{kiyota26}. 
Interestingly, we find that a large number of sources among the extreme-decrement AGN candidates have V-shaped SEDs ($\sim 50/167$), similar to LRDs. 

Most of them differ from high-redshift LRDs in some aspects. The SDSS sources often show strong \nii$\lambda\lambda6548, 6584$ emission, possibly reflecting different metallicities and/or ionization conditions. They may represent ``high-metallicity LRDs'' in a late evolutionary stage from the \nii\ weak LRDs \citep[see also][for other objects/samples that have the V-shaped SEDs in the low-redshift Universe]{ding26, bao26, seyberlich26}. Many show WISE variability in contrast to local LRDs \citep[e.g.,][]{burke26}. In addition, blue UV photometry can be affected by strong emission lines \citep[e.g.,][for a DOGs study]{noboriguchi22} and/or strong UV scattered light from AGNs. In any case, these objects would be valuable low-redshift comparison candidates, but establishing a physical or evolutionary connection with LRDs requires a detailed analysis, which we defer to future work. 

\bibliography{references}{}
\bibliographystyle{aasjournalv7}

\end{document}